\documentclass[%
twocolumn,
amsmath,amssymb,
aps,
prb,
]{revtex4-1}

\usepackage{hyperref}
\usepackage[T1]{fontenc}
\usepackage[utf8]{inputenc}
\usepackage{lmodern} 
\usepackage{amsfonts}
\usepackage{breakurl} 
\usepackage[switch]{lineno} 
\usepackage{graphicx}
\usepackage{epstopdf}
\usepackage{scrextend}
\usepackage{dcolumn}
\usepackage{bm}
\usepackage{textcase}
\usepackage{textcomp}
\usepackage{mathtools}
\usepackage{float}
\usepackage{amsmath}
\usepackage{pifont}
\usepackage{fleqn}
\usepackage{url}
\usepackage{color}

\newcommand{\ceprcoge}{Ce$_{1-x}$Pr$_x$CoGe$_3$}
\newcommand{\cecoge}{CeCoGe$_3$}
\newcommand{\prcoge}{PrCoGe$_3$}
\newcommand{\cehalfcoge}{Ce$_{0.5}$Pr$_{0.5}$CoGe$_3$}

\begin{document}

\title{Influence of Pr substitution on physical properties of Ce$_{1-x}$Pr$_x$CoGe$_3$ system:\\
A combined experimental and first-principles study}

\author{Przemys\l{}aw Skokowski}%
 \email{przemyslaw.skokowski@ifmpan.poznan.pl}
\author{Karol Synoradzki}%
\author{Miros\l{}aw Werwi{\'n}ski}%
\author{Tomasz Toli{\'n}ski}%
 \affiliation{Institute of Molecular Physics, Polish Academy of Sciences, Smoluchowskiego 17, 60-179 Pozna{\'n}, Poland}

\author{Anna Bajorek}
\author{Gra\.zyna Che\l{}kowska}
\affiliation{Institute of Physics, University of Silesia, Uniwersytecka 4, 40-007 Katowice, Poland}


\begin{abstract}
We present the results of our investigations of physical properties for the novel Ce$_{1-x}$Pr$_x$CoGe$_3$ system performed with a~number of experimental methods: magnetic susceptibility, specific heat, electrical resistivity, magnetoresistance, and thermoelectric power. 
Moreover, the electronic structure was studied by means of photoelectron spectroscopy measurements and first-principles calculations.
All investigated compositions of the Ce$_{1-x}$Pr$_x$CoGe$_3$ series crystallize in the tetragonal BaNiSn$_3$-type structure.
The lattice parameters and unit cell volumes decrease with increasing Pr concentration.
On the basis of the measurements taken, a~preliminary magnetic phase diagram was created.
A continuous suppression of the long-range magnetic ordering was observed with increase of Pr concentration.
The critical Pr concentration for magnetic moment ordering was determined from linear extrapolation of the ordering temperature \textit{versus} $x$ to the lowest temperatures ($T = 0$ K) and is equal to about 0.66.
Based on the first-principles calculations we show how the substitution of Pr for Ce affects the electronic structure and magnetic properties of the considered alloys.
Within a single model we take into account the magnetic ordering, fully-relativistic effects, and Hubbard U repulsion on Ce and Pr.
The impact of Hubbard U on the results of calculations is also discussed.
We present the valence-band analysis, Mulliken electronic population analysis, and calculated electronic specific heat coefficients. 
For \cecoge{} it is found that the $++--$ configuration of magnetic moments on Ce is slightly more stable than the $+-+-$ one, 
and also that the calculated value of total magnetic moment on Ce (including spin and orbital parts) is in good agreement with the measurements.
%
%
%
%
\end{abstract}


\maketitle

\section{\label{intr}Introduction}
Most of the interesting properties of the strongly correlated systems are connected with the presence of the $f$-electrons.
Extensive experimental and theoretical efforts focus on studying the unconventional superconductivity and other effects, like for example: the heavy fermion state, deviations from the Fermi liquid behavior, and competition between the RKKY (Ruderman-Kittel-Kasuya-Yosida) and Kondo interactions~\cite{coleman2005quantum, maple2010non, steglich2016foundations, wirth2016exploring, weng2016multiple, smidman2017superconductivity}.
%
%
%
In this work, we will focus on non-centrosymmetric structures that are of great scientific interest, because the lack of inversion symmetry is a key factor in the formation of unconventional superconductivity. 
%
One of widely studied crystal structures is BaNiSn$ _3 $-type (space group $I4mm$, no. 107).
Many of compounds crystallizing in this structure are composited as $RTX_3$, where $R\ -$ rare-earth, $T\ -$ transition metal, and $X\ -$ Si, Ge or Al.
The examples are: Eu$T$Ge$_3$ ($T=$ Co, Ni, Rh, Pd, Ir, Pd)~\cite{bednarchuk2015synthesis}, $R$CoSi$_3$ ($R=$ Pr, Nd, Sm)~\cite{nallamuthu2016magnetic}, and NdCoGe$_3$~\cite{measson2009magnetic}. 
A wide group of compounds are those containing Ce.
The exemplary compounds are: CeRhSi$_3$\cite{muro1998contrasting}, CeIrSi$_3$\cite{muro1998contrasting}, CeRhGe$_3$\cite{muro1998contrasting}, CeIrGe$_3$\cite{muro1998contrasting}, CeCuAl$_3$\cite{klicpera2015neutron}, and CeAuAl$_3$\cite{adroja2015muon}.
%
%

One of the most interesting $RTX_3$ compound is CeCoGe$_3$, with three magnetic phase transitions at $T_{\rm N1} = 21$ K, $T_{\rm N2} = 12$ K, and $T_{\rm N3} = 8$ K~\cite{thamizhavel2005unique}.
For this compound three metamagnetic transitions at $\mu_0 H_{c1} = 0.19$~T, $\mu_0 H_{c2} = 0.84$~T, and $\mu_0 H_{c3}~=~3.0$~T were observed for $H \parallel [001]$~\cite{thamizhavel2005unique}.
The electronic specific heat coefficient value for this compound is $\gamma~=~32$~mJ~mol$^{-1}~$K$^{-2}$. 
The magnetic structure of each of the magnetic phases described by Smidman \textit{et al.} by means of the neutron diffraction experiment~\cite{smidman2013neutron}.
CeCoGe$_3$ compound undergoes superconducting transition under the pressure $p_{sc} = 5.5$~GPa at temperature $T_{sc}~=~0.7$~K~\cite{kawai2007pressure}.
After substitution of Ge with Si the CeCoGe$_{3-x}$Si$_x$ series exhibits a~quantum critical point (QCP) for the critical concentration $x_c=1.5$ ~\cite{krishnamurthy2002non, kanai1999observation}.
Even more complicated evolution of magnetism occurs in the system CeCo$_{1-x}$Fe$_x$Ge$_3$~\cite{de2001phase, skokowski2017magnetoresistance, skokowski2019electronic}, in which the predominant ferro- or antiferromagnetic contributions change as the Fe content increases.
Another intersting compound of with BaNiSn$_3$-type structure is PrCoGe$_3$.
It is a paramagnet with a~low value of the electronic specific heat coefficient $\gamma~=~6.1$~mJ~mol$^{-1}$~K$^{-2}$~\cite{measson2009magnetic}.
Analysis of magnetic susceptibility with the Curie-Weiss law indicated the Pr$^{3+}$ ion state without Co and Ge contributions to the effective magnetic moment~\cite{measson2009magnetic}.
Moreover, theoretical calculations and de Haas van Alphen experiment revealed identical Fermi surface topologies for PrCoGe$_3$ and LaCoGe$_3$ compounds~\cite{kawai2008split}.
The authors suggest that the Pr $f$-electrons do not contribute to the Fermi surface and increase the cyclotron mass, which is for PrCoGe$_3$ nearly twice as large as for LaCoGe$_3$.
More interestingly, considering similarities to LaCoGe$_3$, the two Pr $4f$ electrons might be in a low-spin state caused by strong interactions with CEF.
The metamagnetic transition observed at The magnetic field value of 50~T and 1.3~K changes the crystal electric field (CEF) scheme~\cite{measson2009magnetic}, which indicates possible metaorbital transition, which is predicted, for example, for CeCu$_2$Si$_2$ compound~\cite{pourovskii2014theoretical}.
Therefore, modification of the local surroundings of Pr ions might change their ground state, causing an increase in the hybridization of $4f$ electrons with conduction band.

The aims of this article are to investigate the effects of the substitution in the Ce$_{1-x}$Pr$_x$CoGe$_3$ system on electronic structure and physical properties, and to identify the role of the rare-earth elements in the magnetism of these alloys.
Our experimental efforts are followed by first-principles calculations.
In case of materials containing rare-earth elements, the simplest approaches based on density functional theory (DFT) are often insufficient.
Even the simplest rare-earth materials, like $\alpha$ and $\gamma$ Ce phases, are recently undergoing an in-depth study to determine the optimal approaches leading to the reliable results~\cite{tran_nonmagnetic_2014}.
In our study we will go beyond the general gradient approximation (GGA) and apply the intra-atomic Hubbard U repulsion term (GGA~+~U).
However, the application of GGA~+~U method for spin-polarized Ce systems raises a problem of an emergence of multiple solutions~\cite{shick_ground_2001, tran_nonmagnetic_2014}.
We will discuss how to find the ground state solution anyway and how the value of Hubbard U parameters affects the results.
In contrast to the previous theoretical works on \cecoge{}~\cite{jeong_electronic_2007, skokowski2019electronic}, the new model will include the antiferromagnetic ordering as observed at low temperatures.
We will also present the DFT results for the terminal \prcoge{} composition and for the \cehalfcoge{} alloy containing two types of rare-earth elements.
An important issue that has to be taken into consideration is the CEF on the Pr site in the temperature range of $10-30$~K, which has been observed by many experimental methods~\cite{measson2009magnetic}.
It is known that the distribution of charge around an ion produces an electric field, which is experienced by the $4f$ electrons.
The type of splitting of the ground state by the CEF depends in particular on the type of ion.
It can provide additional contribution to the temperature dependences of physical properties, e.g. specific heat.
Usually, the contribution of Ce ions is observed in the range of $100-300$~K.
For example for the CeCoGe$_3$ compound the CEF energy levels of $\Delta_1 = 220$~K and $\Delta_2 = 315$~K have been determined by the use of the inelastic neutron scattering method~\cite{smidman2013neutron}.
%

%
%
%
%
%
%
%
%
\section{\label{exp} Methods}
\subsection{Experimental details}
The polycrystalline samples were obtained by melting high purity elements several times in an arc furnace to ensure sample homogeneity.
Final mass loss was less than 1\%.
Next, the sample ingots were wrapped in tantalum foil, encapsulated in evacuated quartz tubes, and annealed at $900^\circ {\rm C}$ for 7 days.
The crystal structure of prepared samples was investigated with X-ray diffraction (XRD) measurements performed on X’pert Pro PANalytical device with Cu-K$\alpha$ radiation source at room temperature.
In order to test the physical properties of the Ce$_{1-x}$Pr$_x$CoGe$_3$ system a~number of experimental methods have been used.
As the main measuring device we used a~Quantum Design Physical Property Measurement System (QD PPMS) with appropriate options for specific measurements.
Firstly, the zero field cooling (ZFC) and field cooling (FC) curves of magnetic susceptibility were measured with a~vibrating sample magnetometer (VSM) module in the temperature range $2-300$ K and at applied magnetic field value of 0.1 T.
Hysteresis loops were measured at 2 K at magnetic field values up to 9 T.
In the next step, specific heat measurements were performed in the temperature range $1.9-295$~K without applied magnetic field and in the range of $1.9-40$~K for various magnetic field values.
Further investigations were focused on electrical resistivity and isothermal magnetoresistance, which were measured using the four probe method.
The resistivity measurements were performed without applied magnetic field in the range of $2-300$~K, while magnetoresistance was measured at magnetic field values up to 9~T for temperatures in the range of $2-30$~K.
Thermoelectric power data were collected using four probe method of the thermal transport option (TTO) in the temperature range $2-300$~K.
Finally, the X-ray photoelectron spectroscopy (XPS) measurements were performed with the use of PHI 5700/660 Physical Electronics spectrometer.
The measured spectra were analyzed at room temperature using monochromatized Al K$\alpha$ radiation (1486.6~eV).
The clean surface of samples was obtained by \textit{in-situ} fracturing.
All procedures and measurements were performed in ultrahigh vacuum chamber (UHV) with base pressure of 10$^{-10}$~Torr. 
\subsection{\label{mab} Computational details}

We will present also the results of density functional theory (DFT) calculations.
The models of \cecoge{}, \cehalfcoge{}, and \prcoge{} are investigated using the full-potential local-orbital scheme (FPLO version 18.00-52)~\cite{koepernik_full-potential_1999}.
The FPLO code is one of the DFT implementations characterized by the highest numerical accuracy~\cite{lejaeghere_reproducibility_2016}.
Its precision comes, among the others, from the use of the full-potential approach,
which does not introduce shape approximation to the crystalline potential and to the expansion of the extended states in terms of localized atomic-like basis orbitals~\cite{koepernik_full-potential_1999,eschrig_2._2004}.
The application of the full-potential method is particularly important for the 4$f$-electron systems for which the results are strongly dependent on the quality of the potential~\cite{koepernik_full-potential_1999}.
Another crucial element of our approach is treating the relativistic effects in a full 4-component formalism.
The application of a fully-relativistic (including spin-orbit coupling) approach significantly improves the description of 4$f$ electrons, which are characterized by a high value of spin-orbit coupling.
For the exchange-correlation potential, we choose the generalized-gradient approximation (GGA) in the Perdew-Burke-Ernzerhof form (PBE)~\cite{perdew_generalized_1996}.
For elements with an open 4$f$ shell, like Ce or Pr, 
it is important to further improve the utilised approximation by applying Hubbard U intra-atomic repulsion term to the energy functional, resulting in LSDA~+~U (local spin density approximation) or GGA~+~U method~\cite{ylvisaker_anisotropy_2009}.
In this work, we used the fully-localized limit of the LSDA+U (GGA+U) functional introduced by Czy{\.z}yk and Sawatzky~\cite{czyzyk_local-density_1994}, sometimes referred also as an atomic limit.
The magnitude of Hubbard U repulsion introduced to the Ce~4$f$ and Pr~4$f$ orbitals has been set at 6~eV, which value has been previously calculated for Ce~\cite{anisimov_density-functional_1991}, wheras the $J$ parameter was set to zero.
The issue of choice of the parameter U we will discuss in more detail in Sec.~\ref{ssec.afm_cecoge}.
We have also checked that the effect of the on-site repulsion U on the Co~3$d$ orbitals is weak, so in the case of \ceprcoge{} system, the U$_{3d}$ corrections can be neglected.
A similar first-principles approach has been taken, for example, for fcc Ce~\cite{tran_nonmagnetic_2014} and PrO$_{2}$~\cite{tran_pbe_2008}.
However, some authors argue that for Ce compounds with transition metals (TM) the application of Hubbard U correction for Ce $4f$ electrons is not necessary because as a result of strong hybridization between the Ce $4f$ and TM $3d$ electrons the bonding bands are filled earlier than is the case of localized $4f$ electrons of pure Ce for which one should use the corrections like for example Hubbard U~\cite{sozen2019ab}.

\begin{table}[h]
\centering
\caption{\label{tab_lattice} 
The lattice parameters of \ceprcoge{} compounds used for first-principles calculations.
}
\vspace{2mm}
\def\arraystretch{1.5}%
\begin{tabular}{cccc}

\hline
\hline
composition		& $a$ ({\AA})	& $c$ ({\AA})	& reference\\
\hline
\cecoge{} 		& 4.32 		& 9.835		& \cite{eom1998suppression} \\
\prcoge{} 		& 4.308 	& 9.829		& \cite{kawai2008split} \\
\cehalfcoge{} 	& 4.314 	& 9.832		& interpolated \\
\hline
\hline
\end{tabular}
\end{table}

\begin{figure}[t!]
\centering
\includegraphics[clip, width = 0.9 \columnwidth]{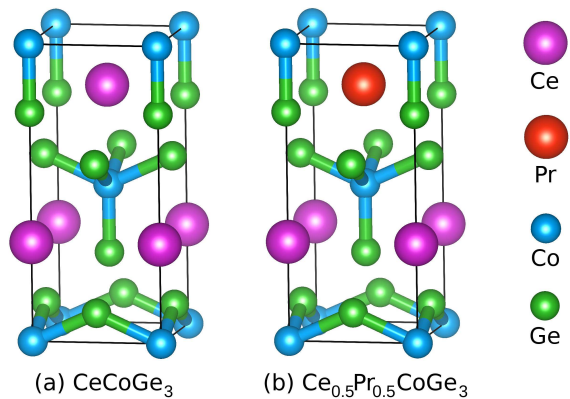}
\caption{\label{fig_struct}The crystal structure models of (a) \cecoge{} and (b) \cehalfcoge{} compositions.
They crystallize in a~tetragonal non-centrosymmetric structure of the BaNiSn$_3$-type, space group $I$4$mm$.
}
\end{figure}

\cecoge{} and \prcoge{} crystallize in a tetragonal non-centrosymmetric structure of the BaNiSn$_3$-type (space group $I$4$mm$), in which Ce and Pr atoms occupy one position, Co atoms another one, and the Ge atoms the two non-equivalent positions~\cite{pecharsky1993unusual,eom1998suppression,kawai2008split}.
In our models, we used the experimental lattice parameters~\cite{eom1998suppression,kawai2008split} and atomic positions~\cite{pecharsky1993unusual,kawai2008split}, see Table~\ref{tab_lattice} and Fig.~\ref{fig_struct}.
The intermediate composition \cehalfcoge{} was modeled as an ordered compound~\cite{eriksson_electronic_1988} with one Ce and one Pr atom per unit cell containing two formula units, see Fig.~\ref{fig_struct}(b).
The antiferromagnetic configurations ($++--$ and $+-+-$) were constructed on the basis of the double unit cells, see Sec.~\ref{ssec.afm_cecoge}.
For calculations based on the single unit cell (non-magnetic solutions) we used k-meshes equal to 20$\times$20$\times$20 and energy convergence criterion equal to 2.72$\times$10$^{-6}$ eV (10$^{-7}$~Hartree).
For calculations based on the double unit cell (magnetic solutions) we used k-meshes equal to 20$\times$20$\times$6 and charge convergence criterion equal to 10$^{-6}$ which simultaneously led to the energy convergence of about 10$^{-6}$ eV or better.
To visualize crystal structures, we used the VESTA code~\cite{momma_vesta_2008}.

Based on the band-structure results we calculated the valence band X-ray photoelectron spectra.
The densities of states (DOS) of individual orbitals have been convoluted by the Gaussian function with a full width at half maximum parameter $\delta$ equal to 0.3 eV.
The aim of convolution was to imitate the experimental broadening resulting from the apparatus resolution, lifetime of the hole states and thermal effects.
Subsequently, the partial DOS were multiplied by the appropriate photoionization cross-sections~\cite{yeh_atomic_1985}.
The above method of determining theoretical X-ray photoelectron spectra we used previously for the Zr-Pd alloys~\cite{skoryna_xps_2016}.
\section{\label{res}Results and discussion}
\subsection{Crystal structure}
\begin{figure}[t!]
\centering
\includegraphics[width = 1.0\columnwidth]{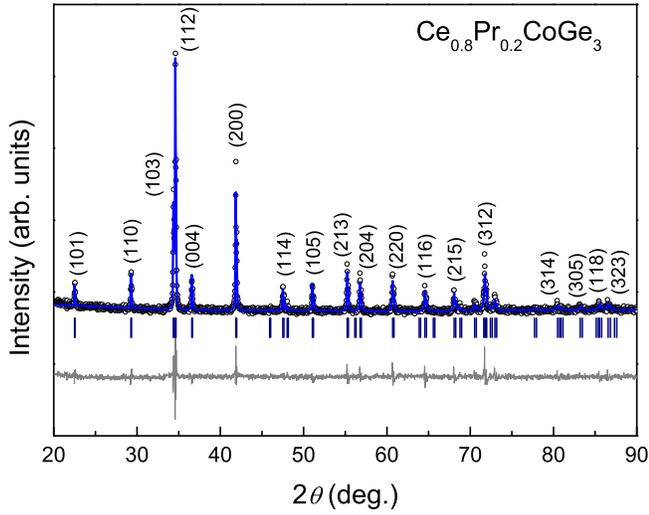}
\caption{\label{fig1}The exemplary X-ray diffraction pattern of the Ce$_{0.8}$Pr$_{0.2}$CoGe$_3$ sample. The bottom solid line shows the difference between the measured and calculated patterns. Vertical bars indicate the positions of structural reflections. Miller indices are presented for the most pronounced peaks.}
\end{figure}

\begin{figure}[t!]
\centering
\includegraphics[width = 1.0\columnwidth]{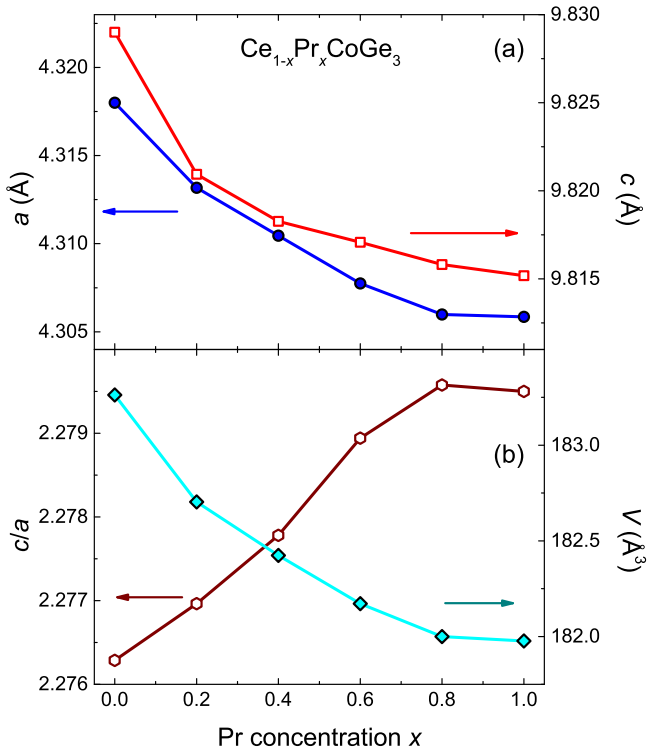}
\caption{\label{fig2}Parameters $a$ and $c$ of the crystal lattice, the primitive cell volume $V$ and the $c/a$ ratio as function of Pr concentration $x$ for the series Ce$_{1-x}$Pr$_x$CoGe$_3$.}
\end{figure}

The crystal structure of the prepared samples was examined by XRD measurements at room temperature.
The results were analyzed using the FULLPROF \cite{rodriguez1993recent} program (an exemplary refinement is shown in Fig.~\ref{fig1}), which revealed that all samples have the desired single-phase non-centrosymmetric tetragonal BaNiSn$_3$-type structure.
The evolution of crystal lattice parameters with Pr concentration is presented in Fig.~\ref{fig2}.
Values of the parameters $a$ and $c$ decrease with the addition of Pr as the alloying ion has smaller radius than the Ce ion.
The $c/a$ ratio presents small growth of values with increasing Pr content.
For the parent compounds, the obtained values of the lattice parameters are in good agreement with those known from the literature~\cite{kawai2008split, eom1998suppression}.

\subsection{\label{magexp}Magnetic properties}
\begin{figure}[t!]
\centering
\includegraphics[width = 1.0\columnwidth]{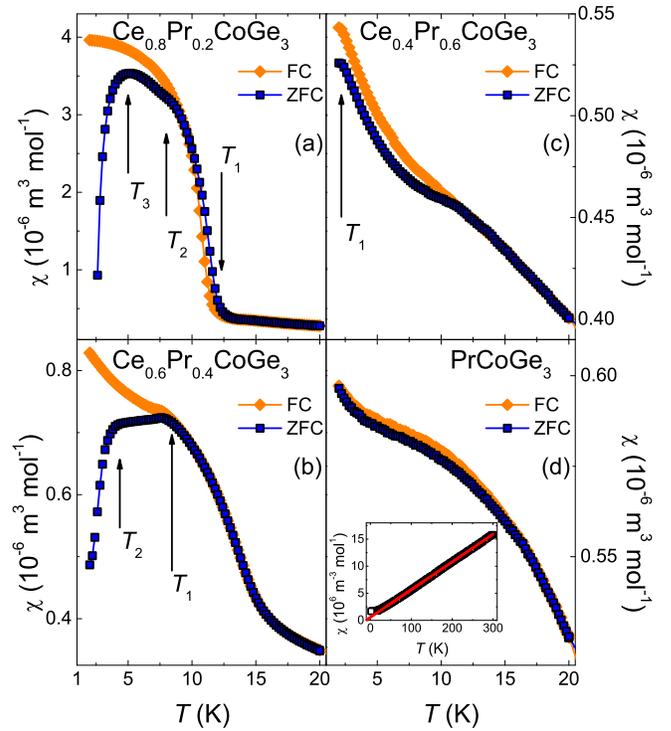}
\caption{\label{fig3}Zero field cooling (ZFC) and field cooling (FC) curves of magnetic susceptibility for the series Ce$_{1-x}$Pr$_x$CoGe$_3$ at $\mu_0 H = 0.1 \ {\rm T}$. Inset of panel (d) shows the inverse magnetic susceptibility fitted with the Curie-Weiss law.}
\end{figure}

\begin{figure}[t!]
\centering
\includegraphics[width = 1.0\columnwidth]{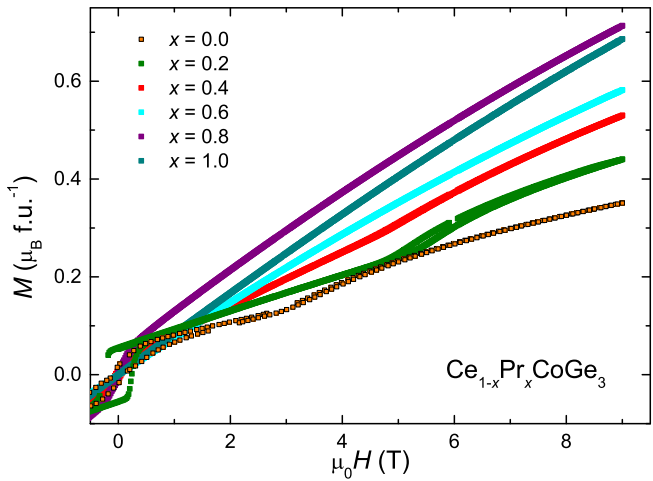}
\caption{\label{fig4}The first quarters of the hysteresis loops for the series Ce$_{1-x}$Pr$_x$CoGe$_3$ at temperature of 2~K.}
\end{figure}

Figure~\ref{fig3} presents temperature dependencies of magnetic susceptibility measured at low temperatures.
In Fig.~\ref{fig3}(a) the results for sample with Pr concentration $x=0.2$ are shown.
We have defined the phase transition temperatures as $T_i$ ($i = 1$, 2, 3), because the nature of the magnetic order varies depending on the extent of substitution.
The magnetic phase transitions for the sample with Pr concentration $x=0.2$ takes the values $T_1 = 11.4(2)$~K, $T_2 = 7.8(2)$~K, and $T_3 = 5.2(2)$~K, which are shifted towards lower temperatures in comparison to the temperatures of magnetic phase transitions of the parent compound CeCoGe$_3$~\cite{thamizhavel2005unique}.
Like for the previously studied CeCo$_{1-x}$Fe$_x$Ge$_3$ system~\cite{skokowski2019comprehensive}, a small amount of the substituted element results in the enhancement of the ferromagnetic correlations.
However, a~metamagnetic transition is still visible in Fig.~\ref{fig4} for the magnetization curves of the samples with Pr content $x=0.2$ and 0.4.
This may suggest that magnetic structure is preserved after the change from antiferromagnetic to ferrimagnetic and only enhancement of the ferromagnetic contribution occurs.
This change might be caused by a~small difference in cell parameters, which changes the RKKY interaction into a~ferromagnetic type or by magnetic disorder, which has a~noticeable contribution visible by the large discrepancy between the ZFC and FC curves.
%
%
In this case, as a magnetic disorder we can interpret disruption of the magnetic structure by non-interactive Pr ions.
%

For the sample with Pr concentration $x=0.4$ the splitting between ZFC and FC curve occurs at $T_1 = 7.6(2)$~K, which can be assigned to the phase transition, whereas the next transition is at $T_2 = 4.8(2)$~K, see Fig.~\ref{fig3}.
A broad peak in the temperature range $10-20$~K is associated with CEF of Pr and it is also visible for samples with $x\geq0.6$.
The splitting between ZFC and FC curves also suggests a~high disorder in the magnetic structure.
In Fig.~\ref{fig4}, for sample with $x=0.4$, there is a~small hysteresis with metamagnetic transition also at the magnetic field value of 5~T.
The sample with Pr content $x = 0.6$ also exhibits splitting between ZFC and FC curves.
A small hump at 2.2(2)~K implies a~possibility of phase transition.
Additionally, for this sample the hysteresis is not visible in Fig.~\ref{fig4}.
Alloy with the Pr concentration $x=0.8$ presents no anomalies in magnetic susceptibility at low temperatures (not shown in this work) and no magnetic hysteresis (Fig.~\ref{fig4}), which indicates possible paramagnetism.
The parent compound PrCoGe$_3$ shows expected paramagnetic behavior with noticeable contribution originating from the CEF excitation (Fig.~\ref{fig3}).
Moreover, it also does not exhibit hysteresis, see Fig.~\ref{fig4}.

\begin{table}[t!]
\caption{\label{tab1}The effective magnetic moment $m_{\rm eff}$, the paramagnetic Curie temperature $\theta_{\rm P}$ and the effective magnetic moment of Pr $m_ {\rm Pr}$ obtained from fitting magnetic susceptibility with the Curie-Weiss law (Eq. \eqref{eqcw}) for the Ce$_{1-x}$Pr$_x$CoGe$_3$ series. Values of $m_ {\rm Pr}$ were calculated with Eq.~\eqref{eqmpr} with assumption of the free Ce$^{3+}$ ion effective magnetic moment value $m_{\rm Ce}$ = $2.54\ \mu_{\rm B}$.}
\centering
\def\arraystretch{1.5}%
\begin{tabular}{cccc}
\hline
\hline
$x_{\rm Pr}$ & $m_{\rm eff}$ ($\mu_{\rm B}$) & $\theta_{\rm P}$ (K) & $m_{\rm Pr}$ ($\mu_{\rm B}$) \\
\hline
0.0 & $2.542(2)$ & $-63.7(2)$ & -\\
0.2 & $2.757(7)$ & $-36.6(6)$ & $3.493(7)$\\
0.4 & $3.035(13)$ & $-24.8(7)$ & $3.653(13)$\\
0.6 & $3.115(6)$ & $-16.2(3)$ & $3.445(6)$\\
0.8 & $3.345(13)$ & $-7.7(4)$ & $3.517(13)$\\
1.0 & $3.519(9)$ & $-10.3(4)$ & $3.519(9)$\\
\hline
\hline
\end{tabular}
\end{table}

The Curie-Weiss law allowed to determine the valence state of Pr in the Ce$_{1-x}$Pr$_x$CoGe$_3$ series.
We used the formula for a magnetic susceptibility temperature dependence:
\begin{equation}
\chi(T)=\chi_0+\frac{N_{\rm A} m_{\rm eff}^2}{3k_{\rm B}(T-\theta_{\rm P})},
{\label{eqcw}}
\end{equation}
where: $\chi_0$ is the temperature independent magnetic susceptibility, $N_{\rm A}$ is the Avogadro's number, $m_{\rm eff}$ is the effective magnetic moment, $k_{\rm B}$ is the Boltzmann constant and $\theta_{\rm P}$ is the paramagnetic Curie temperature.
If we assume that the Co and Ge contribution is negligible, the magnetic susceptibility is a~sum of the Ce and Pr contributions with appropriate proportions depending on the stoichiometry of a~sample.
Hence, to calculate the effective magnetic moment of Pr, $m_ {\rm Pr}$, we used the formula~\cite{tolinski2011spin}:
\begin{equation}
m_{\rm eff}^2=x \cdot m_{\rm Pr}^2+(1-x) \cdot m_{\rm Ce}^2,
{\label{eqmpr}}
\end{equation} 
with assumption of the free Ce$^{3+}$ ion effective magnetic moment value $m_{\rm Ce}$ = $2.54\ \mu_{\rm B}$, according to previous reports that in this crystal structure Ce has the effective magnetic moment close to $2.50\ \mu_{\rm B}$~\cite{pecharsky1993unusual, yamamoto1995cefege, skokowski2019comprehensive}.
The data resulting from the fitting of the Curie-Weiss law are presented in Table~\ref{tab1}.
For all samples, the Pr part of effective moment is around $3.50\ \mu_{\rm B}$, which is close to the Pr$^{3+}$ ion value 3.58~$\mu_{\rm B}$.
The $\theta_{\rm P}$ shows a~tendency to decrease with increasing Pr concentration.
This indicates a~decrease in the collectivity of magnetism and strength of Kondo interaction, which finally leads to paramagnetic state in the parent compound PrCoGe$_3$.

\begin{figure}[t!]
\centering
\includegraphics[width = 1.0\columnwidth]{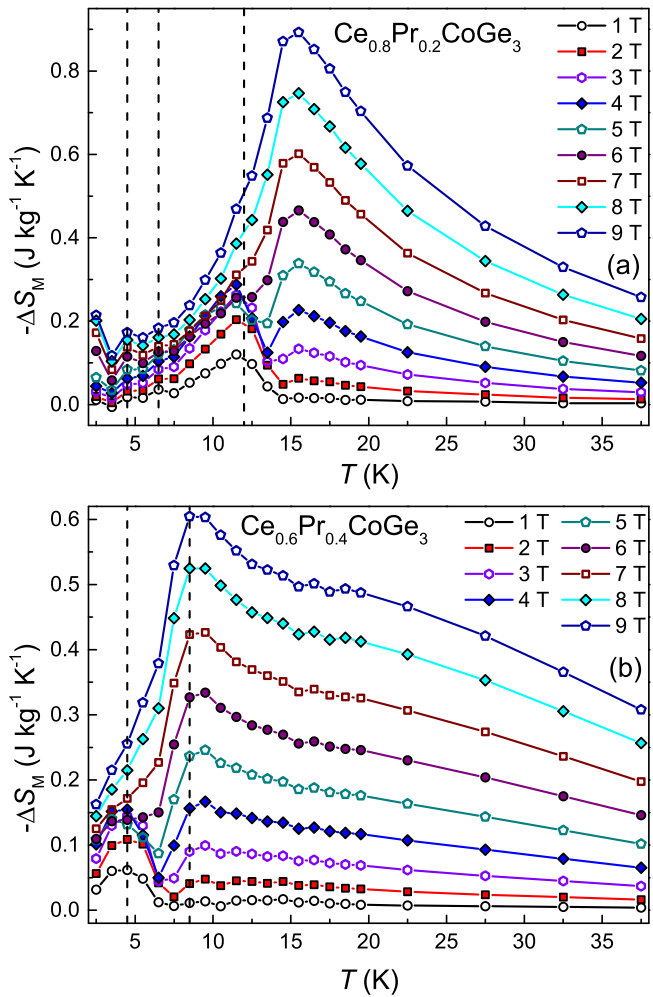}
\caption{\label{figmce}Magnetic entropy change $\Delta S_{\rm M}$ as a~function of temperature $T$ for Pr concentration $x=0.2$ (a) and 0.4 (b) of the Ce$_{1-x}$Pr$_x$CoGe$_3$ series. The vertical lines indicate the detected phase transitions and anomalies.}
\end{figure}

Temperatures of magnetic phase transitions determined from magnetic susceptibility might be inaccurate, because of the overlapping Pr CEF contribution.
Therefore, the phase transition temperatures were additionally estimated employing magnetocaloric effect (MCE).
We calculated the magnetic entropy change $\Delta S_{\rm M}$ with the formula~\cite{gschneidnerjr2005recent}:
\begin{equation}
\begin{aligned}
\Delta S_{\rm M} &\approx \frac{\mu_{0}}{\Delta T}\Big[\int_0^{H_{\max}} M(T+\Delta T, H)\mathrm{d}H \\
&- \int_0^{H_{\max}}M(T,H)\mathrm{d}H\Big],
\end{aligned}
\end{equation}
where: $\mu_0$ is the magnetic permeability of vacuum, $H_{\max}$ is the maximum magnetic field for determined $\Delta S_{\rm M}$, $\Delta T$ is the temperature interval between subsequent isotherms and $M(T, H)$ and $M(T+\Delta T, H)$ correspond to magnetization for specific magnetic field value and temperatures $T$, and $T+\Delta T$. 
Observation of the maxima and minima of $\Delta S_{\rm M}$ can indicate dominating magnetic ordering in the considered samples~\cite{tolinski2012magnetocaloric, skokowski2019comprehensive}.
The plots of $\Delta S_M$ as a function of temperature are presented in Fig.~\ref{figmce}.
%
%
For sample with Pr concentration $x=0.2$ (Fig.~\ref{figmce}(a)) a~wide peak is visible at 11.5(5)~K for the lowest magnetic field values, which corresponds to $T_1$ from the magnetic susceptibility results (Fig.~\ref{fig3}(a)).
Small hump at temperature 6.5(5)~K and small minimum at 4.5(5)~K for magnetic field values of 1~T might correspond to temperatures of phase transitions $T_2$ and $T_3$ denoted from magnetic susceptibility curves.
%
%
For sample with $x = 0.4$ (Fig.~\ref{figmce}(b)) there is also maximum with metamagnetic transition at $T_1 = 8.5(5)$~K, which corresponds to the observed anomaly in the magnetic susceptibility.
A wide peak indicates $T_2 = 4.5(5)$~K.

\begin{figure}[t!]
\centering
\includegraphics[width = 1.0\columnwidth]{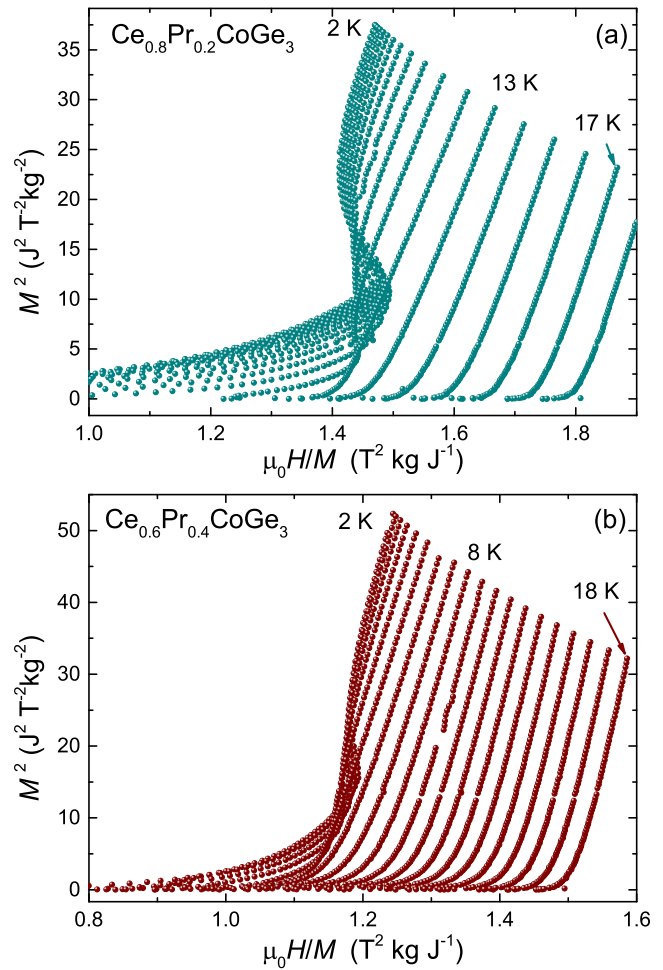}
\caption{\label{figmce2}Arrott plots for samples with Pr concentration $x~=~0.2$ (a) and 0.4 (b) of the Ce$_{1-x}$Pr$_x$CoGe$_3$ series.}
\end{figure}

Arrott plots for samples with Pr concentration $x=0.2$ and 0.4 are presented in Fig.~\ref{figmce2}.
For both samples there is a~negative curvature below the temperature $T_1$. This suggests a~first order metamagnetic transition according to the Banerjee criterion~\cite{banerjee1964generalised}.

\begin{figure}[t!]
\centering
\includegraphics[width = 1.0\columnwidth]{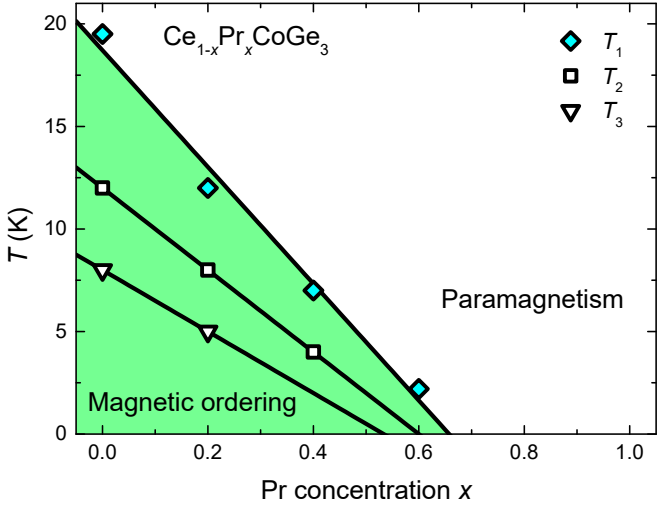}
\caption{\label{fig9}Magnetic phase diagram for the Ce$_{1-x}$Pr$_x$CoGe$_3$ system. Data were obtained from the results of magnetic susceptibility. Solid lines are a~linear extrapolation of specific phase transitions to a~temperature of 0 K. Critical concentrations obtained for specific phase transitions are: $x_{\rm c1}=0.66(8)$ for $T_1$, $x_{\rm c2}=0.60(1)$ for $T_2$, and $x_{\rm c3}=0.53(1)$ for $T_3$. Results for CeCoGe$_3$ are taken from Ref.~\cite{skokowski2019comprehensive}}
\end{figure}

Based on the evolution of the phase transitions temperatures with the Pr concentration $x$ a~magnetic phase diagram is constructed (Fig.~\ref{fig9}).
Linear extrapolation of the magnetic ordering temperature $T_1$ to 0 K gives a~critical Pr concentration equal to $x_{\rm c1} = 0.66(8)$, while for $T_2$ it is equal to $x_{\rm c2}=0.60(1)$, and for $T_3$ it is equal to $x_{\rm c3}=0.53(1)$.
Similarly as in the CeCo$_{1-x}$Fe$_x$Ge$_3$ system~\cite{skokowski2019comprehensive} we observed a~suppression of the magnetism, but the evolution of the magnetic behavior with the concentration of Pr is different.
Alongside with a~decrease of $T_1$, the system exhibits a~decrease in the strength of Kondo interaction.
In terms of the Doniach diagram~\cite{doniach1977kondo}, we can assume that the system is approaching the low energy region of RKKY and Kondo interactions.
It is the opposite situation to the quantum critical point (QCP) region, therefore, in our case, any possible non-Fermi liquid (NFL) behavior between concentration of $x = 0.6$ and 0.8 would be connected with the formation of the Griffiths phase instead of the occurrence of the QCP.
A possible explanation for the reduction of energy of RKKY and Kondo interactions may be related to the interference of both interactions due to the substitution of Pr for Ce in the crystal structure $-$ non-interactive Pr ions disrupt collectiveness of the magnetic structure and also coherence of the Kondo lattice, as the Pr ions separate the interacting Ce ions.
Consequently, in the sample volume, statistically RKKY and Kondo interactions are further reduced with a~higher Pr content.
\subsection{Specific heat}
\begin{figure}[t!]
\centering
\includegraphics[width = 1.0\columnwidth]{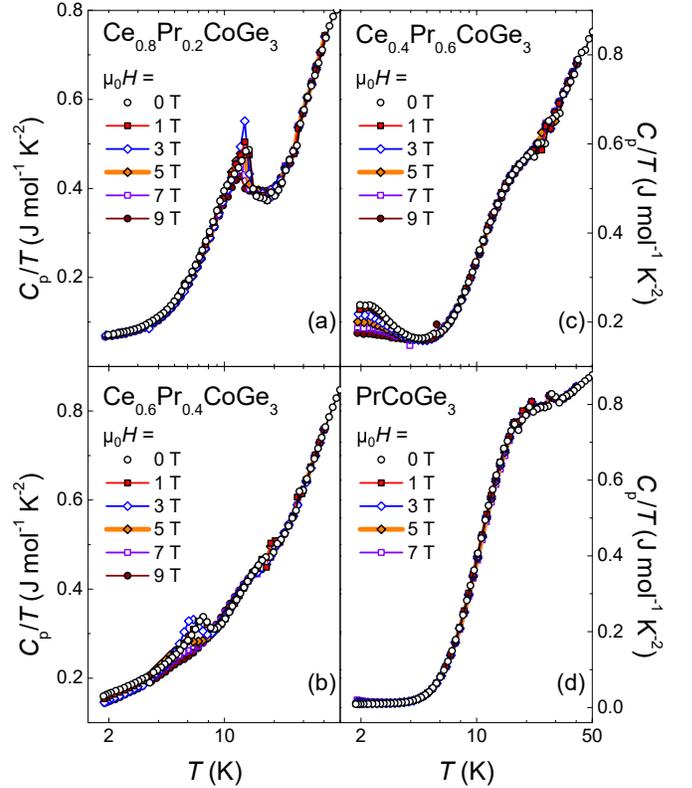}
\caption{\label{fig5}Temperature dependencies of the specific heat of Ce$_{1-x}$Pr$_x$CoGe$_3$ samples presented in the form of $C_{\rm p}/T$ \textit{versus} log$T$ measured in various magnetic field.}
\end{figure}

Temperature dependencies of the specific heat in the range $1.9-40$ K for exemplary samples and different values of the applied magnetic field are presented in Fig.~\ref{fig5}.
%
%
The increase in magnetic field values for samples with Pr concentration $x = 0.2$, 0.4, and 0.6 shifts the main peaks related to the magnetic phase transition $T_1$ towards lower temperatures, which suggests ferrimagnetic or antiferromagnetic type of ordering.
In addition, if we consider the results of magnetic susceptibility and isothermal magnetization measurements, we can assume that the ferrimagnetic ordering scenario is more probable for these samples.

The sample with Pr content $x = 0.4$ presents the suppression of the main phase transition $T_1$ with the increase of the magnetic field values.
For the sample with Pr concentration $x = 0.6$ the transition peak is around 2 K, what can be seen thanks to the influence of magnetic field on the $C_{\rm p}/T$~\textit{versus}~$T$ curves.
Results for PrCoGe$_3$ are similar to those reported earlier~\cite{measson2009magnetic}, with no visible changes due to increased magnetic field values.
For these three samples the CEF contribution connected with Pr is visible as a~wide peak appearing in the temperature range of $10-30$ K.
In addition, the increase in Pr concentration leads to the distinction of the CEF part.
Due to the CEF peak and the magnetic contribution, it is difficult to determine the $\gamma$ parameter values, because the correct temperature range cannot be chosen for applying the formula $C_{\rm p}/T = \gamma + \beta T^2$.
However, we can observe the trend of the $\gamma$ values at 2 K.
For samples with $x = 0.2$ and 0.4 ($x = 0.6$ has a~phase transition at $T = 2.2$ K and it cannot be considered in the analysis) it can be noticed that higher Pr content results in higher $\gamma$ values, which may be related to the increase in disorder.
For parent compounds denoted $\gamma$ values are in good agreement with 32~mJ\,mol$^{-1}$\,K$^{-2}$ for \cecoge{} and 6.1~mJ\,mol$^{-1}$\,K$^{-2}$ for \prcoge{} from previous reports~\cite{thamizhavel2005unique, measson2009magnetic}.
\subsection{Resistivity}
\begin{figure}[t!]
\centering
\includegraphics[width = 1.0\columnwidth]{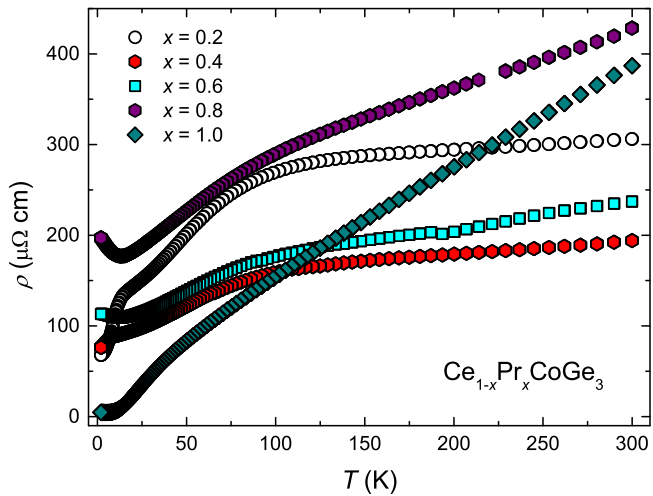}
\caption{\label{fig6}Temperature dependencies of the electrical resistivity for the series Ce$_{1-x}$Pr$_x$CoGe$_3$.}
\end{figure}

The results of the measurements of the temperature dependence of resistivity $\rho$ in the range of $2 -300$ K are shown in Fig.~\ref {fig6}.
With increasing concentration of Pr we can observe a~clear reduction of the Ce CEF contribution (broad peak around 100 K) in favor of the Pr CEF contribution (around 30 K).
This is especially important as we can assume that there are two separate CEF contributions, which are modified by the stoichiometry of the alloy.
It is necessary to consider this information in further investigation of the CEF levels for particular alloys of the system Ce$_{1-x}$Pr$_x$CoGe$_3$.
In Fig.~\ref{fig6} one can also notice anomalies in the low temperature region for samples with Pr content $x = 0.2$, 0.4, and 0.6, which are related to the phase transitions at $T_1$.
For sample with $x = 0.8$ there is an upturn in the lowest temperatures, which might be connected with a~single ion Kondo effect.
This is expected as the concentration of Ce is significantly reduced.
The threshold between the Kondo lattice and the Kondo impurity effect is between $0.6 < x < 0.8$, where also the magnetism is suppressed.
Similarly as for specific heat results, where CEF contribution of Pr and magnetism in the lowest temperatures prevented the systematic estimation of the $\gamma$ parameter, the same issues do not allow to extract the values of residual resistivity $\rho_0$.
However, we can observe some tendencies, which can provide a few rough conclusions.
It is noticeable that the overall $\rho$ values for alloys at the lowest temperatures increase with the addition of Pr, suggesting a~growing chemical disorder (except for the sample with $x = 0.8$ which cannot be considered in the analysis as it presents the Kondo impurity effect).
For the parent compound PrCoGe$_3$, we observe a~metallic type behavior with a~significant CEF contribution in the temperature range of $10-30$~K.
Additionally, the parameter $RRR = 81$ (residual resistivity ratio) of PrCoGe$_3$ has a~rather high value for a~polycrystalline sample and was reproduced for two different pieces of the sample.
Due to the high degree of disorder observed over the whole concentration range (excluding parent compounds), the values of the $RRR$ parameter are only of the order of $1-2$.
%
%
%
%
%
%
\subsection{Magnetoresistance}
\begin{figure}[t!]
\centering
\includegraphics[width = 1.0\columnwidth]{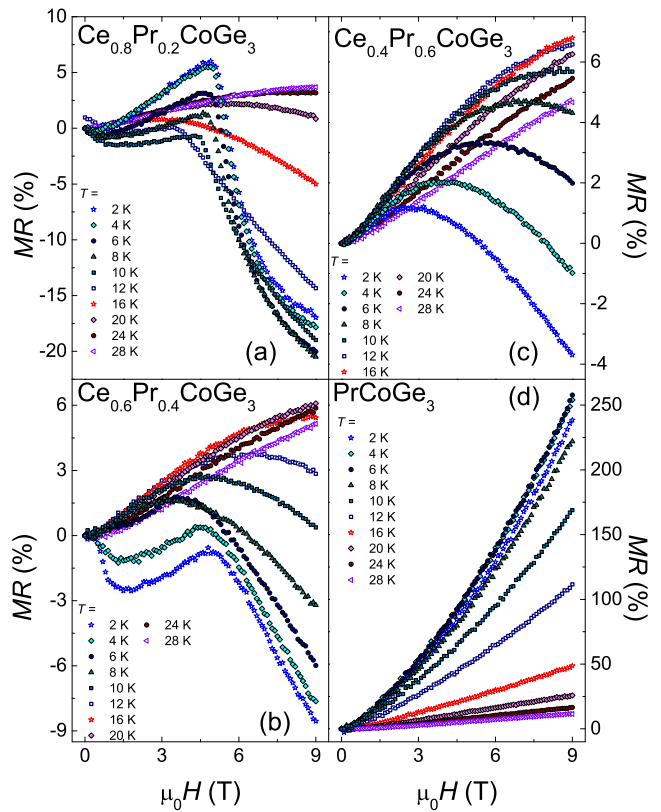}
\caption{\label{fig7}The magnetoresistance ($MR$) isotherms measured in the temperature range $2-30$ K for Pr concentration $ x = 0.2 $, 0.4, 0.6, and 1.0 of the Ce$_{1-x}$Pr$_x$CoGe$_3$ series.}
\end{figure}

In order to extend the characterization of the magnetic properties of the studied alloys, we carried out isothermal measurements of the resistivity as a~function of the magnetic field. 
In Fig.~\ref{fig7} we show magnetoresistance ($MR$) plotted as $MR=[\rho(H,T)-\rho(0,T)]/\rho(0,T)$, where $\rho(0,T)$ and $\rho(H,T)$ correspond to resistivity values for specific temperature without and with applied magnetic field.
For the sample with $x = 0.2$ there is a~metamagnetic transition reflected as a~change from positive to negative values of $MR$ at temperatures below 12 K and at magnetic field of about 5 T.
This confirms the maintenance of the magnetic structure of CeCoGe$_3$ for Ce$_{0.8}$Pr$_{0.2}$CoGe$_3$.
Similar behavior can be observed for the sample with Pr content $x = 0.4$, but the metamagnetic peak is in the region of negative $MR$ values, which may suggest a~ferrimagnetic ordering with a~larger ferromagnetic contribution than in the case of $x = 0.2$.
For higher temperatures, above 18 K for $x = 0.2$, magnetoresistance reveals a~wide positive peak, which can be interpreted as the effect of CEF~\cite{rotundu2007crystalline}.
It is also visible at lower temperatures for samples with $ x = 0.4 $ and 0.6.
Although the CEF contribution increases with the Pr content, it does not coincide with the magnetic part because the magnetic transitions take place at much lower temperatures than the CEF effects.
%
Further increase of temperature causes disappearance of the peak, as we are moving out from the CEF region and a~linear, metalic-type form of the positive $MR$ is revealed.

A more complicated situation is presented in Fig.~\ref{fig7}(d), where the parent compound PrCoGe$_3$ shows a~typical metallic-type $MR$, but with high values of up to 250\%.
While metallic behaviour is expected, the high $MR$ value observed is not so obvious.
%
At first glance, this may be related to the $MR$ formula and low values of $\rho(0,T)$.
Looking at the results for other samples and the contribution of the CEF part to the $MR$, this contribution is not evident in the case of PrCoGe$_3$.
Instead, the CEF part is entirely covered by a~metallic contribution, unlike for sample with Pr concentration $x = 0.2$, where the magnetic contribution overlaps the CEF part.
Previously, the explanation of the giant $MR$ in Tb$_2$Ni$_3$Si$_5$ and Sm$_2$Ni$_3$Si$_5$ has been connected with the layered magnetic structure~\cite{mazumdar1996positive}.
Additionally, for PrNiGe$_3$ compound~\cite{anand2008magnetic}, the authors have suggested the influence of the magnetic field on the mobility of conduction electrons and magnetic ordering.
Since the high values of $MR$ in intermetallic compounds are still not fully explained, we think that in our case the interpretation related to the modification of electron mobility might be possible.
For LaCoGe$_3$~\cite{das2000resistivity}, the results of $MR$ are similar to the results for PrCoGe$_3$.
This is expected because Pr behaves as La in these compounds, providing similar influence on the band structure, as it has been reported by Kawai \textit{et al.}~\cite{kawai2008split}.
This may be an indication that the giant $MR$ for PrCoGe$_3$ compound is associated with the band structure and the mentioned modification of the mobility of the conduction electrons.
\subsection{Seebeck coefficient}
\begin{figure}[t!]
\centering
\includegraphics[width = 1.0\columnwidth]{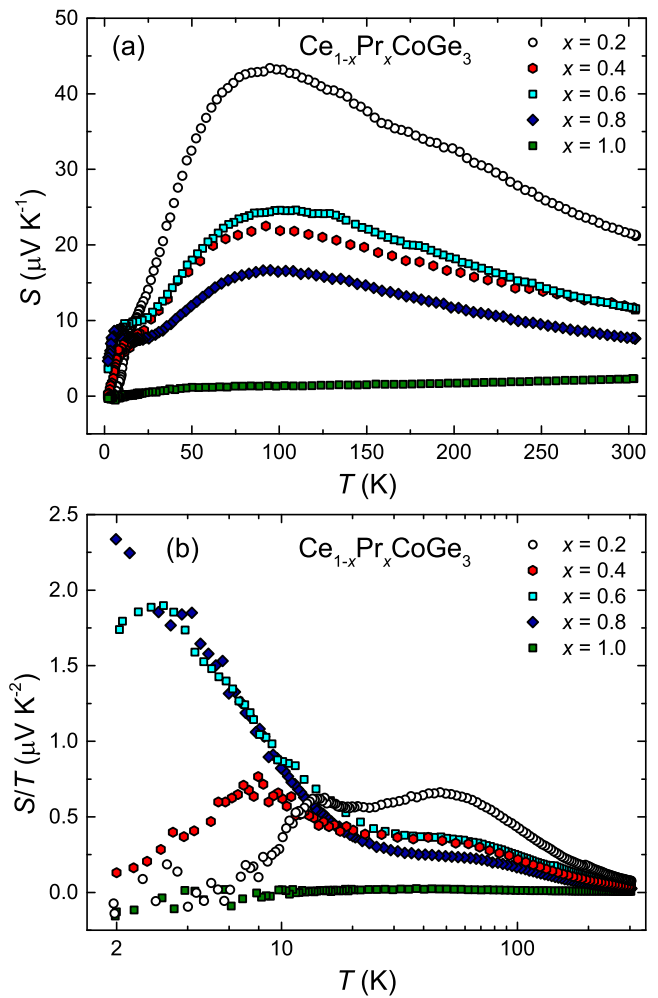}
\caption{\label{fig8} Temperature dependencies of the Seebeck coefficient $S$ of the Ce$_{1-x}$Pr$_x$CoGe$_3$ samples plotted in two different representations: (a) $S$ \textit{versus} $T$ and (b) $S/T$ \textit{versus} log$T$.}
\end{figure}

In Fig.~\ref{fig8}(a) the values of the Seebeck coefficient $S$ as a~function of temperature are presented.
A wide peak around 100 K is observed for all samples containing Ce, which is connected with the Ce CEF contribution.
The highest value $S= 43.4(2)$ $\mu$V~K$^{-1}$ was found for Ce$_{0.8}$Pr$_{0.2}$CoGe$_3$ at 95 K.
Using electrical resistivity values, the thermoelectric power factor $PF=S^2/\rho$ can be calculated.
For sample with $x=0.2$ $PF$ reaches $7.1(1)\times10^{-3}$~W~m$^{-1}$~K$^{-2}$ at 95~K.
The observed decrease in maximum values is associated with the decreasing Ce content, whereas the Pr CEF contribution visible in range of $10-30$ K becomes more pronounced with increasing Pr content.
For PrCoGe$_3$ very low $S$ values are observed.
However, if we consider the formula $S = AT + BT^3$, the linear character of the curve suggests the dominance of the electron diffusion part with almost no contribution of the phonon drag part.
In this case, in low temperature regime the value of one is expected for the Behnia ratio~\cite{behnia2004thermoelectricity}:
\begin{equation}
q=\frac{S}{T} \frac{N_A e}{\gamma}.
\end{equation}
We estimated the $S/T$ values for samples with $x = 0.6$, 0.8, and 1.0 using $C_{\rm p}/T$ values at $T = 2$~K and assuming $q = 1$.
The results are 2.48~$\mu$V~K$^{-2}$, 1.37~$\mu$V~K$^{-2}$, and 0.09~$\mu$V~K$^{-2}$, respectively, which can be compared to the data in Fig.~\ref{fig8}(b).
Good agreement (considering that we do not know exact values of $S/T$ and $C_{\rm p}/T$ at 0~K) between the thermoelectric power and specific heat results confirms the domination of the electronic contribution in low temperature properties of alloys with Pr concentration $x\geq 0.6$.
Therefore, the drop of the $S$ values with increasing $x$ can be connected with decreasing density of electronic states at the Fermi level due to the reduction of the Ce content.

Figure~\ref{fig8}(b) shows the $S/T$ \textit{versus} log$T$ curves for all samples.
Peaks at 12 K, 7 K, and 3 K for Pr content $x$ equal to 0.2, 0.4, and 0.6, respectively, are connected with the magnetic phase transitions.
However, the increase in $S/T$ values with decreasing temperature occurs for the concentration of $x = 0.6$, which is partially overlapped by the phase transition peak.
This is also observed for the Pr concentration of $x = 0.8$, where the Kondo impurity effect is involved.
Observation of this trend may suggest the possibility of NFL behavior for a~very small range of Pr concentration between regions with magnetic ordering and Kondo impurity. However, there are no signs of NFL temperature dependencies in other experimental results. 
\subsection{\label{xpsexp}X-ray photoelectron spectroscopy}
\begin{figure}[t!]
\centering
\includegraphics[width = 1.0\columnwidth]{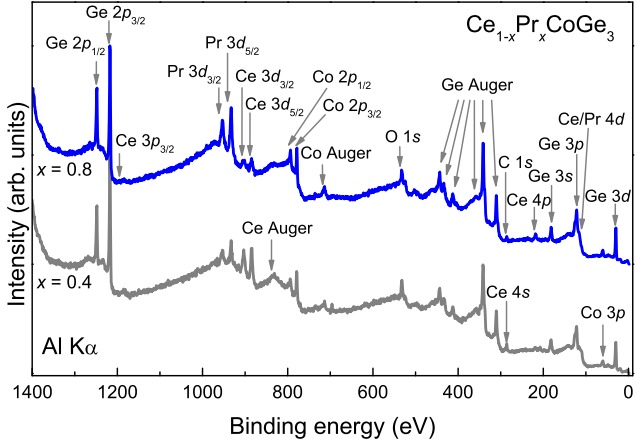}
\caption{\label{figx1}XPS survey spectra of Ce$_{1-x}$Pr$_x$CoGe$_3$ series for selected samples with $x = 0.4$ and 0.8.}
\end{figure}

Exemplary X-ray photoelectron spectra of Ce$_{1-x}$Pr$_x$CoGe$_3$ alloys collected in a~wide binding energy (BE) range up to 1400 eV with the identification of core levels and Auger lines are shown in Fig.~\ref{figx1}.
For all samples we observe low content of oxygen and carbon, suggesting good quality of the samples received.

\begin{figure}[t!]
\centering
\includegraphics[width = 1.0\columnwidth]{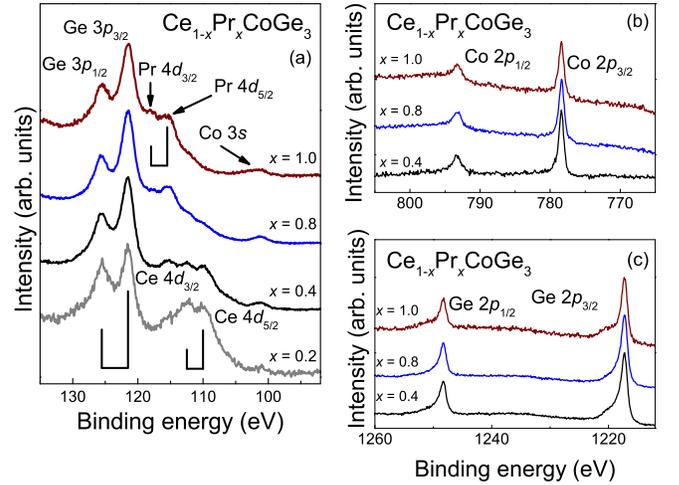}
\caption{\label{figx3}XPS spectra of (a) Ge $3p$, Ce $4d$, and Pr $4d$, (b) Co $2p$, and (c) Ge $2p$ core levels for the Ce$_{1-x}$Pr$_x$CoGe$_3$ samples.}
\end{figure}

In BE region from 90 eV to 140 eV (Fig.~\ref{figx3}(a)) we observed three sets of spin-orbit splitted peaks related to Ce $4d$, Pr $4d$, and Ge $3p$ states.
The positions of the peaks are in good agreement with the ones published for similar systems~\cite{baer1978x, synoradzki2014x, tolinski2017influence} and calculated positions presented in Fig.~\ref{figene}.
Additionally, a~broad peak assigned to Co $3s$ state is present at 101.0 eV.
Due to the overlapping of the Ce $4d$ spectrum with the Pr $4d$ and Ge $3p$ states, the evaluation of the Ce oxidation state is not possible.
While the relative intensity of the peaks associated with Ge is constant for all samples, the intensity of the peaks related to Pr and Ce varies significantly for different samples. 

\begin{figure}[t!]
\centering
\includegraphics[width = 0.9\columnwidth]{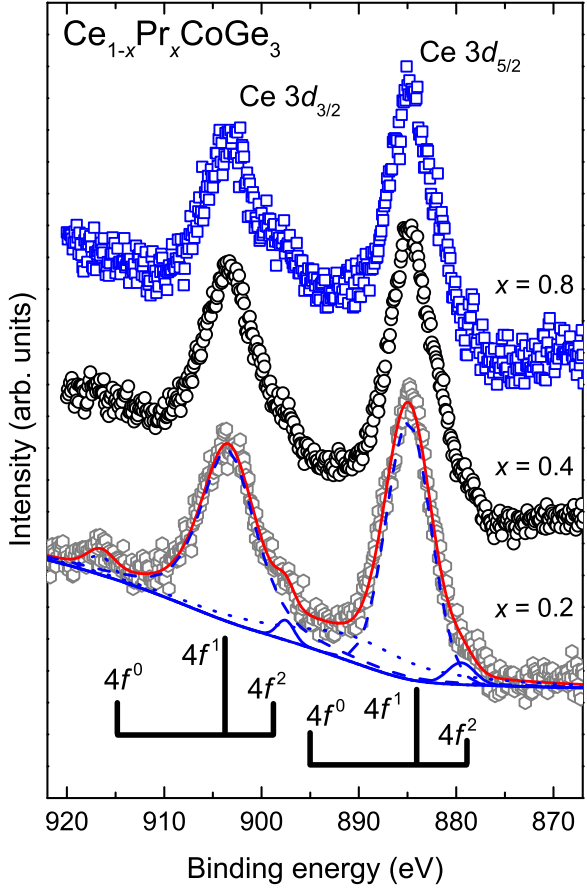}
\caption{\label{figx4}The Ce $3d$ XPS spectra for the selected Ce$_{1-x}$Pr$_x$CoGe$_3$ samples. The spectra intensities are normalized to the intensity of the Ce $3d_{5/2}$ peak. For selected sample, Ce$_{0.8}$Pr$_{0.2}$CoGe$_3$, the deconvolution of Ce $3d$ states is presented (lines) together with levels diagrams. Open symbols correspond to the experimental spectrum.}
\end{figure}

\begin{figure}[t!]
\centering
\includegraphics[width = 0.9\columnwidth]{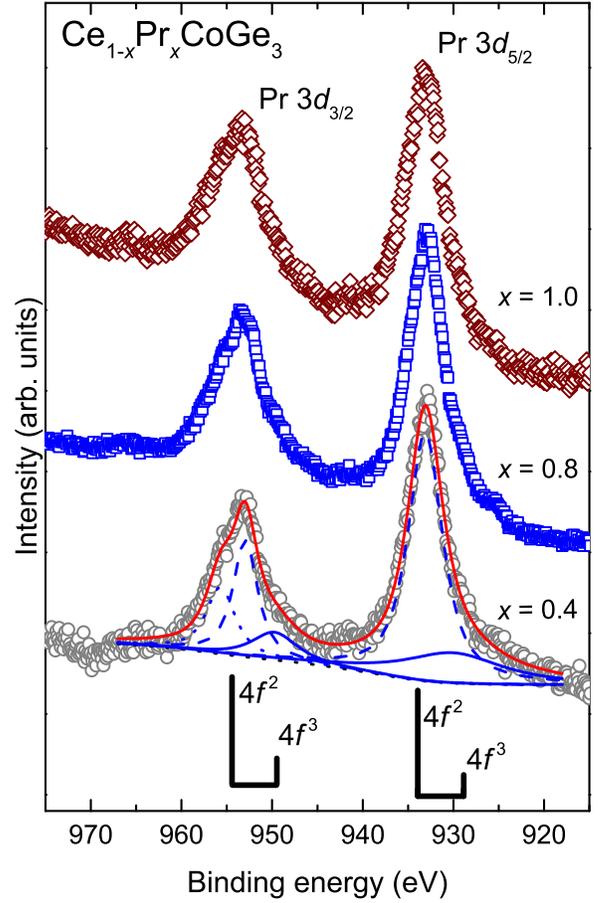}
\caption{\label{figx5}The Pr $3d$ XPS spectra for the selected Ce$_{1-x}$Pr$_x$CoGe$_3$ samples. The spectra intensities are normalized to the intensity of the Pr $3d_{5/2}$ peak. For selected sample, Ce$_{0.6}$Pr$_{0.4}$CoGe$_3$, the deconvolution of Pr $3d$ states is presented (lines) together with levels diagrams. Open symbols correspond to the experimental spectrum.}
\end{figure}

The Co $2p$ XPS spectra for selected samples from Ce$_{1-x}$Pr$_x$CoGe$_3$ series are shown in Fig.~\ref{figx3}(b).
The positions of the Co $2p_{1/2}$ and $2p_{3/2}$ core levels in all investigated samples are around 793.4 and 778.3 eV, respectively, with the spin-orbit splitting value of 15.1 eV.
The values obtained are practically the same as for pure metallic Co~\cite{biesinger2011resolving}.
The lack of additional satellite peaks suggests that there are no Co oxides and that the charge transfer can be neglected in this system.
Therefore, the oxidation state of Co in the Ce$_{1-x}$Pr$_x$CoGe$_3$ series is equal to zero.
Ge shows a~strong peak positions at 1217.3 and 1248.3 eV, as shown in Fig.~\ref{figx3}(c).
The two characteristic peaks correspond to Ge $2p_{1/2}$ and Ge $2p_{3/2}$.
Spin-orbit splitting is equal in this case to 31.0 eV.
In addition to these two sharp peaks, two more wide satellites have been registered towards higher BE values.
These structures probably originate from Ge oxide~\cite{ohta2011x}.
Figures~\ref{figx4} and~\ref{figx5} present $3d$ states of Ce and Pr for selected samples of the Ce$_{1-x}$Pr$_x$CoGe$_3$ series.
In all cases we observed a set of two broad peaks, which originate from spin-orbit splitting of the $3d$ states equal to 18.7~eV and 20.4 eV for Ce and Pr, respectively.
The resulting spectra for Pr look very similar to those published for pure Pr and other compounds containing Pr~\cite{lutkehoff19953d, slebarski1995electronic, slebarski1996crystallographic}.
No visible changes in the Ce and Pr $3d$ spectra have been observed with the change of Pr concentration.
Analysis of XPS spectra for $3d$ states of Ce and Pr with Gunnarsson-Schönhammer theory can provide useful information about valence state and hybridization strength~\cite{gunnarsson1983electron, ogasawara1994lifetime}.
%
%
The exemplary analysis of the Ce $3d$ states for the sample Ce$_{0.8}$Pr$_{0.2}$CoGe$_3$ is presented in Fig.~\ref{figx4}, and for the Pr $3d$ states of Ce$_{0.6}$Pr$_{0.4}$CoGe$_3$ in Fig.~\ref{figx5}.
To model the background, the Tougaard algorithm was used~\cite{tougaard1982influence}.
Relatively small values of the intensity ratio $r_2 = I(f^2)/[I(f^1)+I(f^2)]$, where $I(f^1)$ and $I(f^2)$ are intensities of specific states, suggests weak hybridization of the $4f$ and conduction electrons.
Moreover, according to the ratio $r_0 = I(f^0)/[I(f^0)+I(f^1)+I(f^2)]$, where $I(f^0)$ is the intensity of the $f^0$ state, the absence of distinctive peaks of the Ce $4f^0$ states indicates a~full occupancy of the $f^1$ state.
In order to get the best possible fits, it was necessary to include additional peaks (dotted lines in Fig.~\ref{figx4} and~\ref{figx5}).
In case of Ce $3d$ states we considered two additional peaks at BE~$\approx 892$~eV and BE~$\approx 916$~eV, while for Pr $3d$ spectra additional peak at BE~$\approx 955$~eV was added.
Those additional structures may originate from various excitations, such as plasmon energy loss~\cite{yamasaki2005bulk}.

\begin{figure}[t!]
\centering
\includegraphics[width = 1.0\columnwidth]{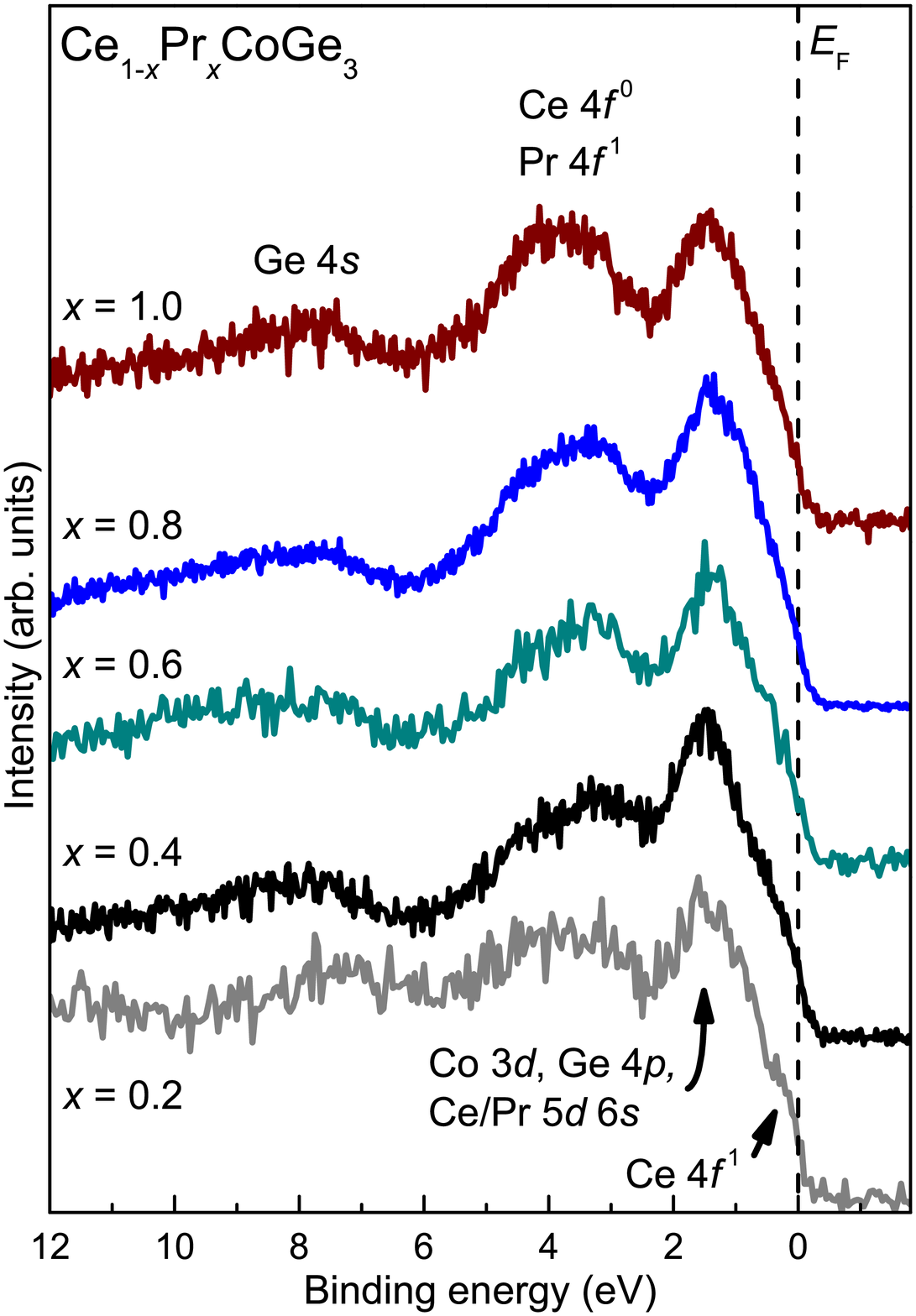}
\caption{\label{figx2}XPS valence band spectra of the Ce$_{1-x}$Pr$_x$CoGe$_3$ series.}
\end{figure}

For all studied samples the XPS valence bands (VBs) in the energy range $0-12$ eV are presented in Fig.~\ref{figx2}.
In this BE range we observed three broad structures.
The broad peak closest to the Fermi level (BE $\approx 2.0$~eV) is formed by the Pr/Ce ($5d$, $6s$), Co $3d$, and Ge $4p$ states.
Small hump close to the Fermi level is related to the Ce $4f^1$ state and its intensity decreases with the increasing Pr concentration.
The middle peak (BE $\approx 4.0$~eV) is mainly due to the Pr $4f^1$ and Ce $4f^0$ states~\cite{penc2007magnetic}.
Therefore, the position and relative intensity of this peak changes with the change of Ce/Pr concentration.
That interpretation is further supported by our DFT results presented in Sec. \ref{abi}.
The last structure, broad band of small intensity near 8.0 eV, originates from Ge $4s$ states.

\subsection{\label{abi}First-principles calculations}

\subsubsection{Relativistic atomic energies}

\begin{figure}[t!]
\centering
\includegraphics[trim = {53 0 0 0}, clip, width = \columnwidth]{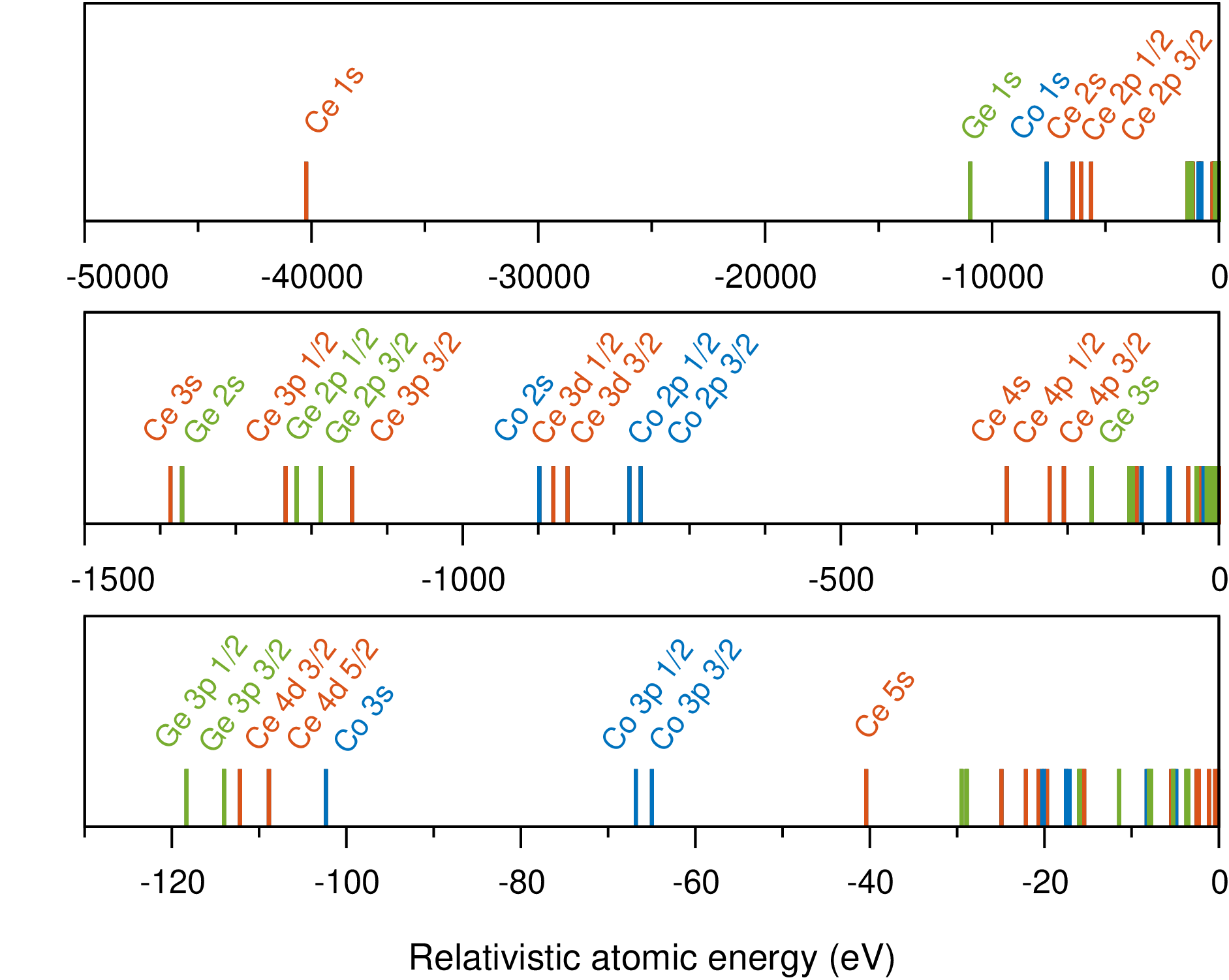}
\caption{\label{figene}The relativistic atomic energies for \cecoge{}
as calculated with FPLO18 in fully-relativistic approach and applying PBE.
}
\end{figure}

We start the theoretical analysis by presenting the electronic structure of CeCoGe$_3$ for a wide range of energies, including core states.
Figure~\ref{figene} shows the relativistic atomic energies in three successively decreasing ranges.
Where the scale allows for it, the atomic energies are attributed to particular orbitals.
The spectrum starts with the strongest bonded electron Ce~1$s$ at about -40~keV.
As the presented atomic energis are calculated in the beggining of the self-consistent cycle, some of the energy levels can shift after the system converge.
Nevertheless, we observe a good agreement between the calculated energy levels presented in the middle panel and the measured XPS survey spectrum shown before, see for example the spin-orbit splitted spectra for Ge~2$p$ (below -1210 eV), Ce~3$d$ (at about -900 eV), and Co~2$p$ (at about -790~eV) in Figs.~\ref{figx3} and \ref{figx4}.
%
The spin-orbit doublets measured for Ge~3$p$ and Ce~4$d$ in a range between -130 and -110~eV, see Fig.~\ref{figx3},
can be identified in the bottom panel of Fig.~\ref{figene}.
However, the XPS spectra closest to the Fermi level are better interpreted on the basis of band structure calculations.

\subsubsection{Valence band X-ray photoelectron spectra}

\begin{figure}[t!]
\centering
\includegraphics[clip, width = \columnwidth]{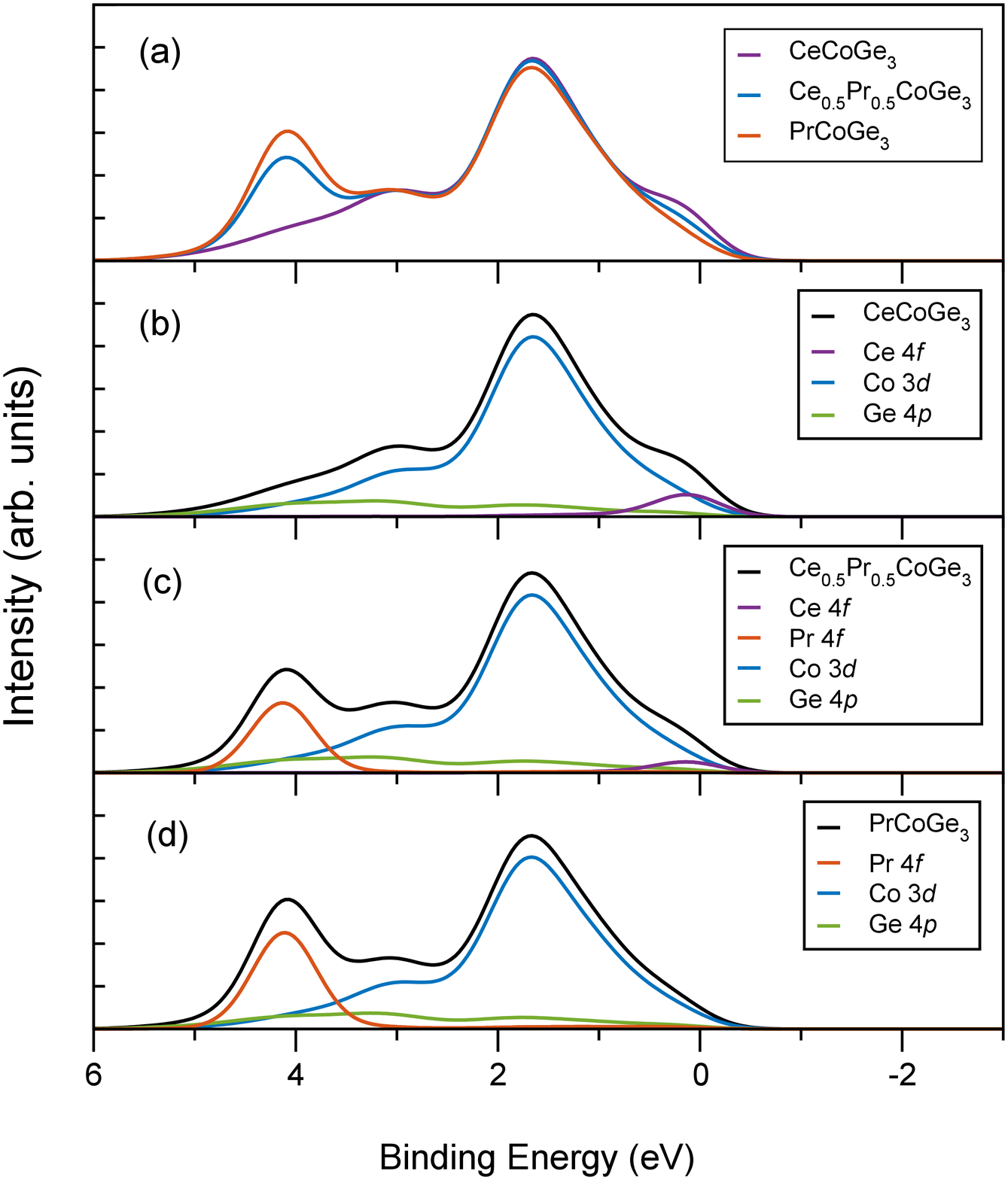}
\caption{\label{fig_xps}The valence band X-ray photoelectron spectra 
calculated with FPLO18 in fully-relativistic approach applying PBE~+~U (U$_{4f}$~=~6~eV)
and assuming lack of spin polarization.
(a) Comparison of the results for \cecoge{}, \cehalfcoge{} and \prcoge{}. 
(b, c, d) The most important contributions from individual orbitals compared to the total XPS spectra.
}
\end{figure}

After a general introduction to the electronic structure of \cecoge{}, we will now focus on a valence band covering a range of only several eVs around the Fermi level.
A detailed analysis of the valence band allows us to predict many physical properties of the materials tested,
whereas the primary purpose of our first-principles calculations is the interpretation of the measured XPS spectra and investigations of magnetic properties.
In our previous work on CeCo$_{1-x}$Fe$_x$Ge$_3$ alloys~\cite{skokowski2019electronic} we presented the XPS spectra calculated without on-site repulsion term U.
Because, in contrast to the \cecoge{} case, the Hubbard U applied to Pr~4$f$ orbitals significantly affects the valence band below Fermi level, in this work we decided to consequently present all the calculated XPS spectra including on-site repulsion term U$_{4f}$ equal to 6~eV, see Fig.~\ref{fig_xps}.

A comparison of the XPS spectra calculated for 
\cecoge{}, \cehalfcoge{}, and \prcoge{} shows that 
the energy bands located between 0 and 6 eV binding energy consist mainly of Co~$3d$ contribution and a much smaller Ge~4$p$ share.
With increasing Pr concentration, the new maximum for Pr~4$f$ is formed and develops at about 4~eV, and at the same time, the Ce~4$f$ contribution observed at Fermi level disappears.
The main features of the calculated XPS spectra are in good agreement with the experimental results shown in Fig.~\ref{figx2}.
%

\subsubsection{Antiferromagnetic solutions for \cecoge{} \label{ssec.afm_cecoge}}

%
%
Our  magnetic measurements for \cecoge{}, see Fig.~\ref{fig9},
confirmed the appearance of three magnetic phase transitions at 21~K, 12~K, and 8~K observed previously~\cite{thamizhavel2005unique}.
Those results have been interpreted as the transitions between paramagnetic, ferrimagnetic, and antiferromagnetic configurations~\cite{ pecharsky1993unusual}.
Pecharsky et al.~\cite{ pecharsky1993unusual} deduced from magnetization isotherms that in temperature below 16~K (a magnetic ground state in their understanding) the Ce magnetic moments are primarily antiferromagnetically ordered in the ab plane, but canted along the c axis. 
The canting order is $+-+-$, i.e., a colinear antiferromagnetic ordering along the c axis~\cite{pecharsky1993unusual}.
The total longitudinal moment at 3~K has been deduced by those authors to be equal to 0.37~$\mu_{\mathrm{B}}$~\cite{pecharsky1993unusual}.
However, the refinement of the integrated intensities of single crystal neutron diffraction, presented by Smidman group, suggests a two-up, two-down magnetic structure ($++--$) below 8~K, with magnetic moments of 0.405~$\mu_{\mathrm{B}}$/Ce atom along the c axis~\cite{smidman2013neutron}.
As the Hund's-rule value of the ground-state magnetic moment ($gJ$) for Ce$^{3+}$ ion ($M_J$~=~5/2, $J$~=~5/2) is 2.14~$\mu_{\mathrm{B}}$, the experimental moment for \cecoge{} is clearly reduced against this theoretical value.
However, such reduction is a well recognized characteristic of the Ce intermetallic compounds, where, for example, the measured magnetic moments on Ce are 
about 0.5~$\mu_{\mathrm{B}}$ for CeFe$_2$~\cite{kennedy_magnetic_1990}, 
1.0~$\mu_{\mathrm{B}}$ for CeB$_6$~\cite{sato_magnetic_1984},
and 1.47~$\mu_{\mathrm{B}}$ for CeSi~\cite{shaheen_observation_1987}.
In addition to measuring the magnetic moment on Ce at 2~K, Smidman's group also determined the CEF scheme for \cecoge{}~\cite{smidman2013neutron}.
The predicted CEF ground state wave function $\psi_1$ ($\Gamma_6 (1)$) corresponds to the Ce magnetic moment of 1.01~$\mu_{\mathrm{B}}$ along the c axis~\cite{smidman2013neutron}.
Smidman and coworkers conclude that the lower value of the measured moment on Ce (0.405 $\mu_{\mathrm{B}}$) compared to the deduced value of the CEF ground state results from a hybridization between the ground state and conduction electrons~\cite{smidman2013neutron}.

%
%
In order to examine the magnetic properties of \cecoge{}, we decided to extend the non-spin-polarized first-principles investigations into the spin-polarized procedure.

Since CeCoGe$_3$ in its ground state is an antiferromagnet with a non-trivial configuration of magnetic moments extending beyond a single elementary cell, it is necessary to pay special attention to preparing a suitable model.
At the same time, in case of describing the Ce 4$f$ electrons it is necessary to go beyond the LDA/GGA.
The procedure for preparing such a model is not standard and therefore it is described in the Appendix.
We show there how we have dealt with the problem of emergence of multiple solutions within the proposed model combining the LDA/GGA~+~U approach with description of the magnetic properties of Ce systems.
We discuss the influence of the value of the Hubbard U parameter on the obtained values of magnetic moments and the densities of electronic states (DOS) of the valence band and
we conclude that the results are not very sensitive to the value of U in a range from about 3 to 6~eV.
Finally, in Appendix we present the models of the most probable collinear antiferromagnetic configurations ($++--$ and $+-+-$) deduced from experiments.
We find that the $++--$ configuration oriented along the [100] direction is the ground state and the $+-+-$ [100] one is slightly less stable.
In all cases, the spin magnetic moment on Ce is close to 1.00~$\mu_{\mathrm{B}}$, which is related to a localization of the occupied 4$f$ orbitals.
The utilized fully-relativistic approach allows us to calculate also the orbital contributions to the magnetic moments, which in the case of $f$-electron systems are often substantial~\cite{morkowski_x-ray_2011}.
In our case, the opposite orbital magnetic moment on Ce is equal to about -0.57~$\mu_{\mathrm{B}}$ and significantly reduces the resultant total magnetic moment equal to about 0.43~$\mu_{\mathrm{B}}$.
The latter result stays in a good agreement with the experimental value 0.405~$\mu_{\mathrm{B}}$/Ce deduced from single crystal neutron diffraction below 8~K~\cite{smidman2013neutron} and with another experimental result suggesting the magnetic moment of about 0.37~$\mu_{\mathrm{B}} $/Ce at 3~K~\cite{ pecharsky1993unusual}.
Additionally, the calculated moments on Co are below 0.04~$\mu_{\mathrm{B}}$/atom and are opposite to the moments on Ce, whereas the moments on Ge are below 0.005~$\mu_{\mathrm{B}}$/atom and parallel with Ce.

%
%
\begin{figure}[t!]
\centering
\includegraphics[clip, width = \columnwidth]{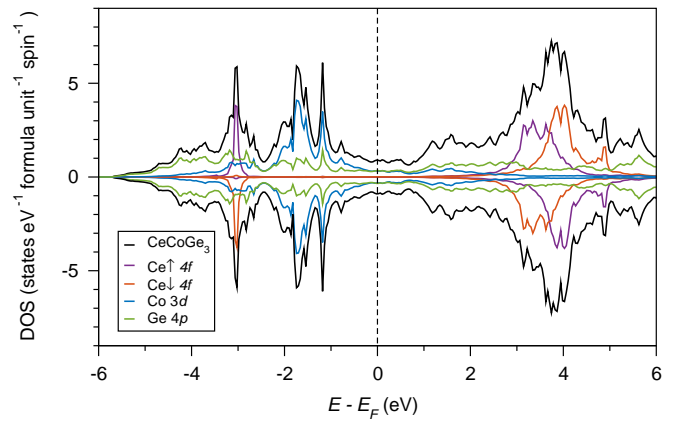}
\caption{\label{fig_dos_afm}The densities of states (DOS) of \cecoge{}.
The model is based on a double unit cell with antiparallel configuration of magnetic moments on Ce sites ($++--$) and quantization axis [100]. 
The total DOS is presented together with the most significant contributions of individual orbitals (Ce 4$f$, Co 3$d$, and Ge 4$p$).
The results are obtained 
with FPLO18 in fully-relativistic approach and applying PBE~+~U (U$_{4f}$~=~6~eV).
}
\end{figure}
Figure \ref{fig_dos_afm} shows the spin-polarized DOS for \cecoge{} with the ground-state antiparallel configuration $++--$ [100] of magnetic moments on Ce sites, whereas the magnetic configuration sketch is presented in Fig.~\ref{fig_mag_stuct}(f) of Appendix.
The observed hybridization between the Co~3$d$ and Ge~$4p$ orbitals in a whole region below the Fermi level confirms the formation of the 3$d$-4$p$ covalent bonds. 
As we have already discussed, the position of 4$f$ bands depends on the selected value of the Hubbard U.
Hence, for U$_{4f}$ equal to 6 eV the occupied 4$f$ states form the lower Hubbard band at about -3~eV and upper Hubbard band centered at about 4~eV, which stay in decent agreement with experimental values for $\gamma$ Ce (-2 and 4 eV~\cite{wuilloud_electronic_1983, wieliczka_high-resolution_1984}).
The lower Hubbard band is not hybridized with other bands, which indicates localization of the occupied Ce 4$f$ orbital.
What is characteristic to the antiferromagnetic solution, the result consists of two types of Ce contributions with an antiparallel orientation of magnetic moments, which are denoted as Ce~$\uparrow$ and Ce~$\downarrow$ and marked in different colors on the plot.
These two results are symmetric, and each consists of majority spin channel occupied with about one electron (1.09 from Mulliken analysis) and another channel nearly empty (about 0.09~$e$).
This polarization of 4$f$ band leads to the local spin magnetic moment on Ce equal to 1.00~$\mu_{\mathrm{B}}$, see also Table~\ref{tab_mm_mag_configs} of Appendix.
The main difference between the presented non-magnetic and antiferromagnetic solutions is a shift of the 4$f$ occupied band from the Fermi level to the position of -3~eV below Fermi level, compare with Fig.~\ref{fig_xps}(b).

%
The density of states at the Fermi level 
equal to 1.8~states\,eV$^{-1}$\,f.u.$^{-1}$ consists mainly of contributions from Co~3$d$ and Ge~$4p$ orbitals forming the valence band.
This rather low value corresponds to the electronic specific heat coefficient $\gamma$ equal to 4.3~mJ\,mol$^{-1}$\,K$^{-2}$
in qualitative agreement with $\gamma$ equal to 32~mJ\,mol$^{-1}$\,K$^{-2}$ measured for \cecoge{}~\cite{thamizhavel2005unique}.

%
The excess electron number (resultant charge) of \cecoge{} calculated for magnetic configuration $++--$ [100] is the same as obtained for a non-magnetic case.
Just like in the non-magnetic case, the charge taken from Ce sites (-1.23) is transferred to the Co (+0.17) and Ge (+0.47, +0.30) sites.
However, the occupation of particular Ce orbitals is slightly different.
While for non-magnetic solution we had occupation Ce 5$p^{5.82}$ 6$s^{0.19}$ 5$d^{1.59}$ 6$p^{0.14}$ 4$f^{0.98}$, for configuration $++--$ [100] we have Ce 5$p^{5.85}$ 6$s^{0.18}$ 5$d^{1.49}$ 6$p^{0.13}$ 4$f^{1.09}$, 
which differs primarily by a charge 0.1 transferred from Ce~5$d$ to Ce~4$f$ orbital.

\subsubsection{Antiferromagnetic solution for \cehalfcoge{} \label{ssec.afm_ce05pr05coge}}
%
%
%
%
\begin{figure}[t!]
\centering
\includegraphics[clip, width = \columnwidth]{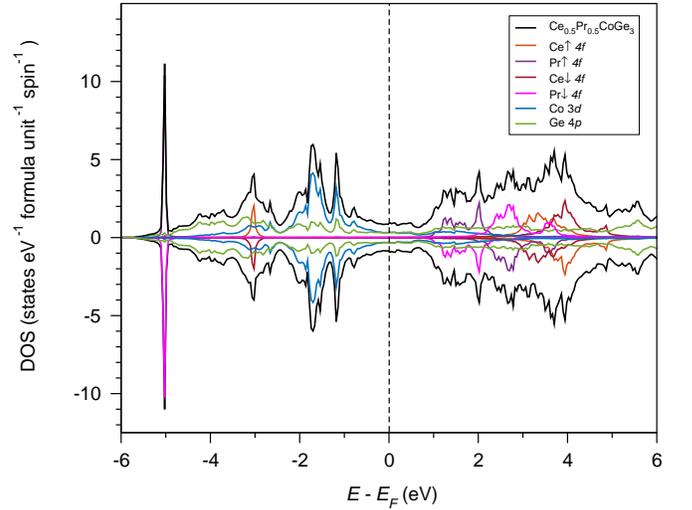}
\caption{\label{fig_ceprcoge_dos_afm}The densities of states (DOS) of \cehalfcoge{} ordered compound.
The model is based on a double unit cell with the antiparallel configuration of magnetic moments Ce(+) Pr(+) Ce(-) Pr(-) along the $c$ axis and with the quantization axis [001] (in the previous section it was [100] that was energetically preferred). 
The total DOS is presented together with the most significant contributions of individual orbitals (Ce 4$f$, Pr 4$f$, Co 3$d$, and Ge 4$p$).
The results are obtained 
with FPLO18 in fully-relativistic approach and applying PBE~+~U (U$_{4f}$~=~6~eV on Ce and Pr sites).
}
\end{figure}
For the composition \cehalfcoge{} we do not perform as extended analysis as for \cecoge{}.
Instead, we present a model based on a double unit cell, with the antiparallel configuration of magnetic moments Ce(+) Pr(+) Ce(-) Pr(-) along the $c$ axis, and with the energetically preferred quantization axis [001].
The DOS of \cehalfcoge{} and \cecoge{} are very similar, see Figs.~\ref{fig_dos_afm} and \ref{fig_ceprcoge_dos_afm}.
The main difference between the occupied parts below the Fermi level is the contribution of Pr~4$f$ states located at about -5~eV.
Additionally, in \cehalfcoge{} the Ce~4$f$ are depopulated due to the lower concentration of Ce in the alloy.
The Mulliken analysis shows that the charges taken from Ce (-1.21) and Pr (-1.14) sites are transferred to Co (+0.18, +0.16) and Ge (+0.47, +0.43, +0.28) sites.
The occupation of Pr valence orbitals 
(5$p^{5.74}$ 6$s^{0.25}$ 5$d^{1.50}$ 6$p^{0.23}$ 4$f^{2.14}$) 
is most different from the Ce counterparts
(5$p^{5.85}$ 6$s^{0.18}$ 5$d^{1.49}$ 6$p^{0.13}$ 4$f^{1.09}$)
for 4$f$ orbital, 
which for Pr is close to two instead of one for Ce.
The occupation of spin-polarized Pr~4$f$ orbital is reflected as a high peak in the DOS and also in the Pr spin magnetic moment equal to 2.08~$\mu_{\mathrm{B}}$.
Taking into account an orbital moment equal to -1.45~$\mu_{\mathrm{B}}$ leads to the total magnetic moment on Pr equal to 0.63~$\mu_{\mathrm{B}}$, 
while the calculated magnetic moments on Ce sites stay the same as for \cecoge{} ($m_s = 1.00$ and $m_l = 0.56 \mu_{\mathrm{B}}$).
Similar like for the antiferromagnetic solution of \cecoge{}, 
the calculated value of DOS at the Fermi level for \cehalfcoge{} is equal to 1.8~states\,eV$^{-1}$\,f.u.$^{-1}$, 
which corresponds to the $\gamma$ coefficient equal to 4.3~mJ\,mol$^{-1}$\,K$^{-2}$, 
with regard to the $\gamma$ equal to 32 and 6.1~mJ\,mol$^{-1}$\,K$^{-2}$ measured for \cecoge{} and \prcoge{}, respectively~\cite{thamizhavel2005unique, measson2009magnetic}.

\subsubsection{Non-magnetic solution for \prcoge{} \label{ssec.afm_prcoge}}
%
%
%
%
\begin{figure}[t!]
\centering
\includegraphics[clip, width = \columnwidth]{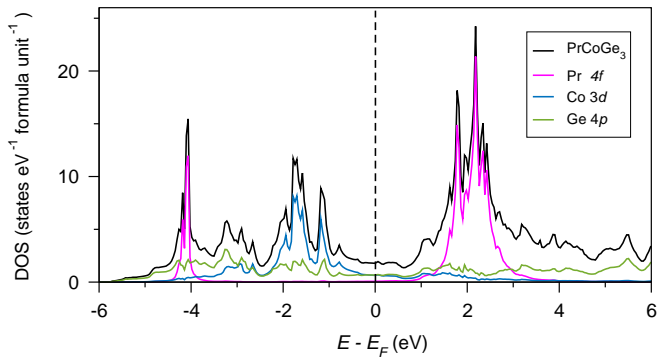}
\caption{\label{fig_prcoge_dos}The densities of states (DOS) of \prcoge{} in the non-magnetic state.
The total DOS is presented together with the most significant contributions of individual orbitals (Pr 4$f$, Co 3$d$, and Ge 4$p$).
The results are obtained 
with FPLO18 in fully-relativistic approach and applying PBE~+~U (U$_{4f}$~=~6~eV).
}
\end{figure}

Because \prcoge{} does not order magnetically, we decided to present only the non-magnetic DOS for this system, see Fig.~\ref{fig_prcoge_dos}, which we previously used to get the XPS spectra, see Fig.~\ref{fig_xps}(d).
We notice that even in the non-magnetic state the occupied Pr~4$f$ states form the lower Hubbard band at about -4~eV below the Fermi level, when for the non-magnetic solution of \cecoge{} we observed much wider 4$f$ peak located directly at Fermi level~\cite{skokowski2019electronic}.
The main contributions of Co~3$d$ and Ge~4$p$ orbitals do not change qualitatively in comparison to the previously discussed \cecoge{} and \cehalfcoge{} results.
Like it was discussed before, the charge taken from Pr (-1.14) sites is transferred to the Co (+0.18) and Ge (+0.45; +0.26) sites.
The occupations of particular Pr orbitals are 5$p^{5.74}$ 6$s^{0.25}$ 5$d^{1.51}$ 6$p^{0.23}$ 4$f^{2.15}$, with regard to the ground state electronic configuration of a neutral Pr atom ([Xe] 4$f^{3}$ 6$s^{2}$).
Similar like for the antiferromagnetic solutions of \cecoge{} and \cehalfcoge{}, 
the DOS at the Fermi level for \prcoge{} is equal to 1.8~states\,eV$^{-1}$\,f.u.$^{-1}$, 
which corresponds to the $\gamma$ coefficient equal to 4.3~mJ\,mol$^{-1}$\,K$^{-2}$,
in fair agreement with the $\gamma$ measured for \prcoge{} equal to 6.1~mJ\,mol$^{-1}$\,K$^{-2}$~\cite{measson2009magnetic}.

\subsubsection{Summary of the first-principles calculations}

Investigating the \ceprcoge{} system (where $x~=~0.0$, 0.5, and 1.0) by first-principles calculations we observed the evolution of the electronic structure with increasing Pr concentration.
The growth of the number of electrons in the system 
due to a higher proportion of the element with a higher atomic number
(Z$_{\rm Ce}$~=~58 and Z$_{\rm Pr}$~=~59)
is realized mainly by increasing the occupation of the narrow 4$f$ band located several eVs below the Fermi level, 
remaining the part of the valence band in the vicinity of the Fermi level almost intact.
Moreover, even though we are able to suggest the ground state ordered magnetic configuration for \cecoge{}, the applied methods are insufficient to model the antiferromagnetic-paramagnetic transition and to indicate the critical concentration.
For example, the paramagnetic state of \prcoge{} could be better investigated with the disordered local moments approach (DLM) based on the coherent potential approximation (CPA), but in FPLO this option is incompatible with the fully-relativistic PBE~+~U approach.

\section{Conclusions}
We have presented results of our studies of various physical properties for the Ce$_{1-x}$Pr$_x$CoGe$_3$ system, which shows the suppression of magnetic ordering with increasing Pr concentration $x$ in isostructural transformation.
From the magnetic hysteresis and magnetoresistance measurements we deduced that the magnetic structure of the parent compound CeCoGe$_3$ is preserved for a~low content of Pr ($x<0.4$).
In all presented experimental results, there is a~significant crystal electric field contribution of Pr in the temperature range of $10-30$~K, which is separated from a Ce contribution visible in the temperature of 100~K.
The Pr CEF contribution may have a~strong impact on the Pr state.
In the studied alloys Pr behaves as La with two additional non-interactive $f$-electrons.
%
%
More interestingly, the parent compound PrCoGe$_3$ shows giant magnetoresistance, which is probably connected with the observed reduction of electrons mobility with the applied magnetic field.
The constructed magnetic phase diagram reveals that the ordering temperature decreases monotonically with increasing Pr concentration $x$ and the extrapolation to 0~K provides the critical concentration $x_{\rm c1}=0.66(8)$ for the phase transition at $T_1$.
The possibility of appearance of the non-Fermi liquid behavior is very low, as the quantum critical point scenario is excluded due to the reduction of energy of RKKY and Kondo interactions.
X-ray photoelectron spectroscopy measurements confirmed the successful synthesis of good quality samples of solid solutions for the Ce$_{1-x}$Pr$_x$CoGe$_3$ series.
The Co $2p$ spectra demonstrate the absence of contribution of Co oxides and a~close similarity with metallic Co.
The analysis of the Pr/Ce $3d$ and $4d$ XPS spectra suggests weak hybridization between the Pr/Ce $4f$ and remaining valence band electrons in the studied materials.
%
The stable Pr$^{3+}$ and Ce$^{3+}$ ions are consistent with magnetic susceptibility measurements.
From the point of view of possible applications, the materials tested, like most of similar compounds based on Ce, show small values of magnetic entropy change (for $x =0.2$ $-\Delta S_{\rm M} = 0.9$~J~kg$^{-1}$~K$^{-1}$ at 15~K for 9~T) and thermoelectric power factor (for $x = 0.2$ $PF = 7.1(1)\times10^{-3}$~W~m$^{-1}$~K$^{-2}$ at 95~K), which does not augur well for their wide application in industry.
We have also presented the results of first-principles calculations for \cecoge{}, \cehalfcoge{}, and \prcoge{} compositions.
We considered non-spin-polarized models as the basis for XPS spectra and spin-polarized models to investigate magnetic properties.
We have shown that the substitution of Pr for Ce significantly affects the electronic structure of the alloys.
The main conclusions of the calculations are as follows:

(1) The calculations explain the evolution of the measured \ceprcoge{} XPS spectra mainly as a change in the position and magnitude of the 4$f$ contributions. 

(2) Mulliken electronic population analysis indicates electronic states of $f$-electron atoms as close to Ce~$f^1$ and Pr $f^2$.

(3) The charge analysis and calculated densities of states indicate that the chemical bonds in \ceprcoge{} alloys are formed mainly by the Ce/Pr 5$d$, Co 3$d$, Ge 4$p$, and Ge 4$s$ orbitals.

(4) The calculated electronic specific heat coefficient $\gamma$ is low and 
remains almost constant as the Pr concentration increases.

(5) The charge transfer mainly occurs from the 4$f$ elements (Ce and Pr) towards Ge and Co.

%
(6) Of the considered magnetic configurations of \cecoge{}, the more stable are the ones with [100] magnetization direction. 

(7) The most stable configuration of \cecoge{} is the antiferromagnetic $++--$, but it has only slightly lower energy than the $+-+-$ one.

(8) The selection of the on-site repulsion term U$_{4f}$ in the range from about 3 to 6~eV does not change qualitatively the calculated densities of states for antiferromagnetic \cecoge{}.

(9) The calculated total magnetic moment on Ce (0.43~$\mu_{\mathrm{B}}$) for \cecoge{} is consistent with the experimental values determined at low temperatures.

%
%
%
%
%
%
%
%
\section*{Acknowledgments}

MW acknowledges the financial support of the National Science Centre Poland under the decision DEC-2018/30/E/ST3/00267.
Part of the computations was performed on the resources provided by the Pozna{\'n} Supercomputing and Networking Center (PSNC).
We thank Dr. Justyna Rychły, Dr. Jan Rusz, and Wojciech Marciniak for reading the manuscript and helpful discussion.

\section*{Appendix A: Mulliken electronic population analysis}

\begin{table}[ht!]
\centering
\caption{\label{tab_charge} 
Excess electron number (resultant charge) for \cecoge{} and \prcoge{} compounds
calculated with FPLO18 in fully-relativistic approach applying PBE~+~U (U$_{4f}$~=~6~eV) and without spin polarization.
The non-equivalent crystallographic sites are based on the Refs.~\cite{pecharsky1993unusual,kawai2008split}.
Unlike the other sites, the multiplicity for Ge2 per formula unit is two instead of one.
\vspace{2mm}
}
\def\arraystretch{1.5}%
\begin{tabular}{ccccc}
\hline
\hline
formula$\backslash$site& Ce/Pr & Co & Ge1 & Ge2 \\
\hline
CeCoGe$_3$ 	&  -1.23 & 0.17 & 0.47 & 0.30 \\
PrCoGe$_3$ 	&  -1.16 & 0.18 & 0.43 & 0.27 \\ 
\hline
\hline
\end{tabular}
\end{table}

\begin{table}[ht!]
\centering
\caption{\label{tab_mulliken} 
The Mulliken electronic population analysis for \cecoge{} and \prcoge{} compounds
calculated with FPLO18 in fully-relativistic approach applying PBE~+~U (U$_{4f}$~=~6~eV) and without spin polarization.
The non-equivalent crystallographic sites are based on the Refs.~\cite{ pecharsky1993unusual,kawai2008split}.
Results for several almost empty valence orbitals were not provided, although they were included in the calculation.
\vspace{2mm}
}
\def\arraystretch{1.5}%
\begin{tabular}{ccccccc}
\hline
\hline
           & site& 5$p$  &  6$s$ &  5$d$ &  6$p$ &  4$f$\\
\hline           
CeCoGe$_3$ &  Ce &  5.82 &  0.19 &  1.59 &  0.14 &  0.98\\
PrCoGe$_3$ &  Pr &  5.73 &  0.25 &  1.50 &  0.22 &  2.15\\
\hline
          &  site& 4$s$  &  3$d$ &  4$d$ &  4$p$\\
\hline          
CeCoGe$_3$ &  Co &  0.52 &  7.91 &  0.19 &  0.53\\
PrCoGe$_3$ &  Co &  0.52 &  7.92 &  0.19 &  0.54\\
\hline
           &  site& 4$s$ &  4$p$ & 4$d$\\
\hline           
CeCoGe$_3$ &  Ge1 & 1.60 &  2.71 & 0.16\\ 
PrCoGe$_3$ &  Ge1 & 1.59 &  2.53 & 0.17\\ 
\hline
           &  site& 4$s$ &  4$p$ & 4$d$\\
\hline           
CeCoGe$_3$ &  Ge2 & 1.60 &  2.55 & 0.17\\
PrCoGe$_3$ &  Ge2 & 1.60 &  2.53 & 0.17\\
\hline
\hline
\end{tabular}
\end{table}

The analysis of the valence band can be extended by an estimation of the partial atomic charges based on the Mulliken approach~\cite{mulliken1955electronic}.
Tables~\ref{tab_charge} and \ref{tab_mulliken} present the results of the Mulliken electronic population analysis for \cecoge{} and \prcoge{}.
The Mulliken approach is possible to apply as the utilized FPLO code is based on the linear combination of atomic orbitals method.
%
%
For \cecoge{}, we observe that the charge taken from Ce sites (-1.23) is transferred to the Co (+0.17) and Ge (+0.47, +0.30) sites.
(A similar picture we see also for \prcoge{}.)
%
%
For both compounds the chemical bonds result mainly from the interaction of Ce/Pr 5$d$, Co 3$d$, and Ge 4$p$ and 4$s$ orbitals.
%
%
For Ce and Pr sites we observe a slight depopulation of 5$p$ orbitals and a low occupation of 6$s$ and 6$p$ orbitals.
A significant difference between the Ce and Pr electronic configurations is found for 4$f$ orbitals (0.98 \textit{versus} 2.15) and stems directly from the distinction in atomic numbers of the elements under consideration (Z$_{\rm Ce}$~=~58 and Z$_{\rm Pr}$~=~59).
The calculated occupations of Ce/Pr 4$f$ orbitals ($f^{0.98}$ and $f^{2.15}$) are in fair agreement with the main contributions identified from analysis of Ce 3$d$ and Pr 3$d$ spectra coming from $f^1$ and $f^2$ states, see Figs.~\ref{figx4} and~\ref{figx5}.
%
%
In the case of Co sites, with regard to the ground state electronic configuration of a neutral Co atom (3$d^7$ 4$s^2$), we observe depopulation of the 4$s$ orbitals and increase in the occupation of the 3$d$ orbital.
In addition, the Co 4$d$ and 4$p$ orbitals become partially occupied.
The latter so-called polarization states are not taken into account in the basic electronic configurations of the neutral atoms or ions.
%
%
In the case of Ge sites, with regard to the ground state electronic configuration of a neutral Ge atom (3$d^{10}$ 4$s^2$ 4$p^2$), we observe accumulation of charge on 4$p$ and 4$d$ orbitals and a partial depopulation of 4$s$ orbitals.

\section*{Appendix B: Details of antiferromagnetic solutions for C\lowercase{e}C\lowercase{o}G\lowercase{e}$_3$}

Tran et al. have summarized that the theoretical approaches most often used to investigate the prototype $\alpha$ and $\gamma$ Ce phases are LDA/GGA, LDA/GGA~+~U, self-interaction corrected LDA (SIC-LDA), and LDA plus dynamic mean-field theory (LDA + DMFT), for more details see Ref.~\cite{tran_nonmagnetic_2014}. 
As it is well known that the LDA/GGA alone does not reproduce the properties of Ce compounds well, we will go beyond the GGA and apply the intra-atomic Hubbard U repulsion term (GGA~+~U), which is often used to model $f$-electron systems.
We choose the U$_{4f}$ to be equal to 6~eV -- a the value previously calculated for Ce~\cite{anisimov_density-functional_1991}.
Furthermore, we take into account the antiferromagnetic configurations $+-+-$ and $++--$ suggested by the experiments~\cite{ pecharsky1993unusual, smidman2013neutron}.
However, before we proceed with the presentation of the results, we would like to discuss the issue of multiple magnetic solutions and the impact of the choice of the value of U$_{4f}$ on the results obtained.

%
%
Among many difficulties related to the LDA/GGA~+~U description of 
the magnetic properties of Ce systems, 
the most important seems to be the emergence of multiple solutions~\cite{shick_ground_2001, tran_nonmagnetic_2014}.
If the applied potential is orbital-dependent, as for the LDA/GGA~+~U functional,
the resultant occupation of $f$ orbitals strongly depends on the occupation used to run the self-consistent calculations~\cite{tran_nonmagnetic_2014}.
Different solutions can also be obtained starting from different parameters, like for example the value of initial spin splitting, the value of U, or starting the calculation with the spin-orbit coupling enabled.
In our calculations for antiferromagnetic ($++--$) configuration of \cecoge{}, 
we have found several distinct solutions characterized by different values of magnetic moments.
The three characteristic solutions were: (1) highly magnetic (spin magnetic moment $m_s$ on Ce atoms equal to about 0.97~$\mu_{\mathrm{B}}$ and orbital magnetic moment $m_l$ on Ce atoms equal to about -1.96~$\mu_{\mathrm{B}}$), (2) medium magnetic ($m_s$~=~1.00~$\mu_{\mathrm{B}}$; $m_l$~=~-0.57~$\mu_{\mathrm{B}}$), and non magnetic ($m_s$~=~$m_l$~=~0).
The comparison of total energies indicated the ground state solution with the orbital magnetic moment $m_l$ equal to about -0.57~$\mu_{\mathrm{B}}$.

%
%
The Hartree-Fock interaction energy and the double-counting term of
the LDA/GGA~+~U functionals depend on the occupation matrix $n_{m,m'}$, where $m$, $m'$ means orbital quantum numbers, for orbital $f$ ($l$ = 3) taking 7 values ($2 \cdot l$ + 1) from -3 to 3~\cite{tran_nonmagnetic_2014}.
The occupation matrix can be used, in some approaches even in a fully controlled manner, to define the occupation of given sets of $d$- or $f$-orbitas on given sites~\cite{allen_occupation_2014}.
The occupation matrix of majority-spin Ce 4$f$ orbital of our solutions with $m_l \approx -0.57\ \mu_{\mathrm{B}}$ is:

\begin{center}
$
\begin{bmatrix*}[r]
 0.01 & -0.04 &  0.01 &  0.00 & -0.01 &  0.04 &  0.00 \\       
-0.04 &  0.48 & -0.06 &  0.00 &  0.06 & -0.48 &  0.04 \\     
 0.01 & -0.06 &  0.01 &  0.00 & -0.01 &  0.06 & -0.01 \\   
 0.00 &  0.00 &  0.00 &  0.01 &  0.00 &  0.00 &  0.00 \\         
-0.01 &  0.06 & -0.01 &  0.00 &  0.01 & -0.06 &  0.01 \\         
 0.04 & -0.48 &  0.06 &  0.00 & -0.06 &  0.48 & -0.04 \\         
 0.00 &  0.04 & -0.01 &  0.00 &  0.01 & -0.04 &  0.01 
\end{bmatrix*},
$
\end{center}
whereas the minority-spin states of Ce 4$f$ occupation matrix are nearly empty.
A roughly equal amount of $m_l$~=~2 and -2 in this state corresponds 
(approximately since the solution is fully relativistic and includes spin-orbit coupling) 
to the orbital composed as a superposition of spherical harmonics:
$f_{xyz} = (Y^2_3 - Y^{-2}_3)/(i\sqrt{2})$.
Our solution also correlates to the one called \textit{FM1} obtained for ferromagnetic fcc Ce within the PBE~+~U approach (U~=~4.3~eV)~\cite{tran_nonmagnetic_2014}
characterized by a very similar magnetic state [$m_s$~=~1.2~$\mu_{\mathrm{B}}$; $m_l$~=~-0.5~$\mu_{\mathrm{B}}$ (inside muffin-tin sphere)].
It is worth noting that the compared solutions were obtained using two DFT implementations based on different basis sets, namely local orbitals in the case of FPLO and plane waves in the case of WIEN2k.
The magnetic solutions (ferro- and antiferromagnetic) discussed further in this section are very similar to the one which has been just described.

%
%
%
The next issue we will discuss before presenting the results of spin-polarized calculations is an impact of the value of the Hubbard U parameter.
For this purpose, we will use a representative model of \cecoge{} with antiparallel configuration $++--$ of magnetic moments on Ce sites.
Other considered magnetic configurations will be discussed in details later.
While it is known that the application of on-site Hubbard repulsion term to the Ce~4$f$ orbitals qualitatively change the results of the LDA/PBE calculations for Ce-based systems~\cite{shick_ground_2001, tran_nonmagnetic_2014}, a selection of a specific value of U for a given system is sometimes problematic.
We decided to set the value of U equal to 6~eV as calculated for Ce within DFT~\cite{anisimov_density-functional_1991}.
However, for example, Tran et al. used for bcc Ce the value of U equal to 4.3~eV, which is the average of two values of U calculated with constrained RPA method for $\alpha$ and $\gamma$ Ce phases~\cite{nilsson_ab_2013}.
To better understand how our results are affected by the magnitude of U value, we solve the set of cases with U equal from 0 to 6 eV.
\begin{figure}[t!]
\centering
\includegraphics[clip, width = \columnwidth]{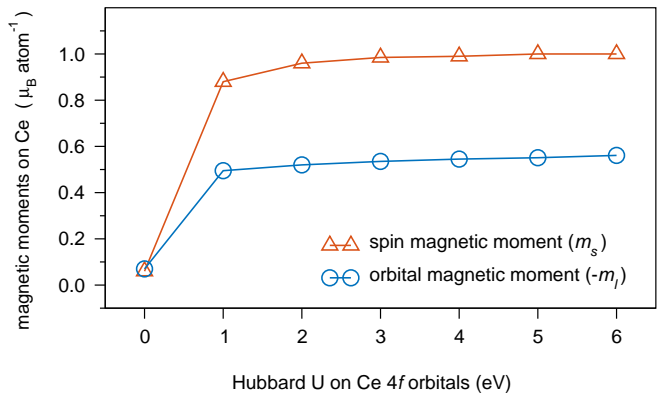}
\caption{\label{fig_mm_vs_u}Spin and orbital magnetic moment on Ce sites as a function of on-site repulsion U$_{4f}$
calculated for \cecoge{} with antiparallel configuration $++--$ of magnetic moments on Ce sites. 
Calculations are carried out 
with FPLO18 in fully-relativistic approach applying PBE~+~U and for quantization axis [001].
While orbital moments are oriented against spin moments, the graph shows orbital moments multiplied by minus one ($-m_l$).
For U~=~0 spin and orbital magnetic moments are equal to 0.06 and -0.07~$\mu_{\mathrm{B}}$, respectively.
}
\end{figure}
%
%
In Fig.~\ref{fig_mm_vs_u} we see that the value of magnetic moment on Ce changes significantly after applying on-site Hubbard repulsion to the Ce~4$f$ orbitals.
At the same time we see that the results are not very sensitive to the value of U in a range from about 3 to 6 eV.

\begin{figure}[t!]
\centering
\includegraphics[clip, width = \columnwidth]{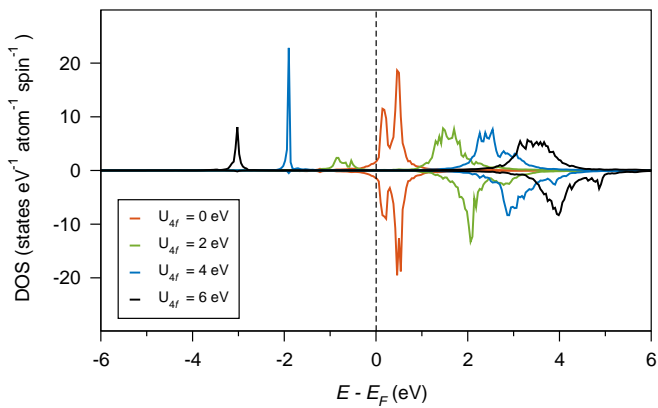}
\caption{\label{fig_dos_ce_vs_u}The densities of states of Ce$\uparrow$~4$f$ orbitals as a function of on-site repulsion U$_{4f}$
calculated for \cecoge{} with antiparallel configuration $++--$ of magnetic moments on Ce sites. 
Calculations are carried out 
with FPLO18 in fully-relativistic approach applying PBE~+~U and for quantization axis [001].
}
\end{figure}
The antiferromagnetic solution for \cecoge{} consists of two types of Ce contributions with antiparallel orientation of magnetic moments denoted as Ce~$\uparrow$ and Ce~$\downarrow$.
Fig.~\ref{fig_dos_ce_vs_u} shows how the value of U affects the densities of states of Ce$\uparrow$~4$f$ orbitals, compare with Ref.~\cite{shick_ground_2001}.
For U$_{4f}$~=~0 the 4$f$ band is located at Fermi level, similar like in non-magnetic case with U$_{4f}$~=~6~eV, see Fig.~\ref{fig_xps}(b).
With an increase in the value of U, the 4$f$ band splits, and the distance between occupied and unoccupied parts grows.
For U~=~6~eV the lower Hubbard band is located at about -3~eV, while the upper band is at about 4~eV above the Fermi level.
For comparison, a photoelectron spectroscopy of Ce $\gamma$ phase indicates lower and upper Hubbard bands at -2 and 4 eV, respectively~\cite{wuilloud_electronic_1983, wieliczka_high-resolution_1984}.
The analysis carried out helps to better understand the effect of U on the results and 
suggests that 6~eV is one of the reasonable values of U$_{4f}$ for Ce.

%
%
As our DFT results present in principle the ground state at 0~K, we are interested in proper modeling of the ground-state magnetic configuration. 
%
%
In the case of \cecoge{}, low-temperature measurements suggest antiferromagnetic ground state.
However, as we have already presented, there is no agreement on the details of the magnetic configuration.
Pecharsky et al. suggest $+-+-$ ordering~\cite{ pecharsky1993unusual}, whereas Smidman et al. deduced $++--$ configuration at 2~K~\cite{smidman2013neutron}.
The next question to which the calculations can provide answers is the direction of local magnetic moments on Ce, however here our method is limited to consider only colinear solutions.
\begin{figure*}[!t]
\centering
\includegraphics[clip, width = 0.95 \textwidth]{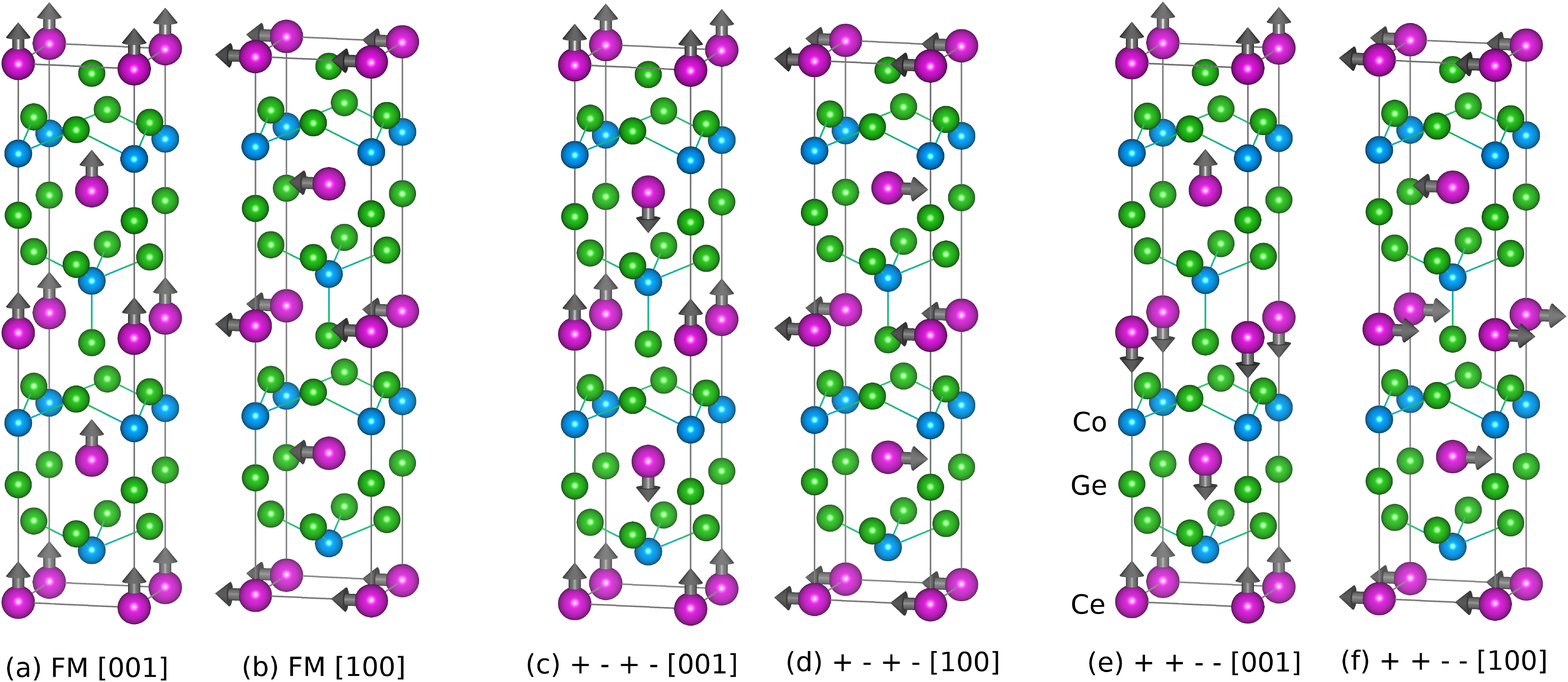}
\caption{\label{fig_mag_stuct}The colinear magnetic configurations of \cecoge{} based on the double unit cell model.
(a, b) ferromagnetic, (c, d) antiferromagnetic $+-+-$, and (e, f) antiferromagnetic $++--$  configurations of spin magnetic moments on Ce sites.
Full relativistic calculations will allow the quantization axes ([001] an [100]) to be considered.
}
\end{figure*}
Looking for an answer, we prepared the models with antiferromagnetic ($+-+-$ and $++--$) and ferromagnetic configurations of spin magnetic moments on Ce sites, see Fig.~\ref{fig_mag_stuct}.
To construct the $++--$ configuration we had to double the unit cell and reduce the symmetry according to magnetic ordering.
To be able to accurately compare the total energies between the  $++--$ configuration and other solutions, the remaining models ($+-+-$ and ferromagnetic) have been prepared within the same double cell approach, even though in these cases in principle it would be possible to use a single unit cell.
In addition, we consider cases with magnetization pointing towards [001] and [100] direction.

%
%
\begin{table}
\caption{\label{tab_mm_mag_configs} 
Spin ($m_\mathrm{s}$) and orbital ($m_\mathrm{l}$) magnetic moments ($\mu_{\mathrm{B}}$ (atom or f.u.)$^{-1}$) on Ce sites calculated for considered colinear magnetic configurations of \cecoge{}, see Fig.~\ref{fig_mag_stuct}.
Magnetic moments are 
calculated with FPLO18 in fully-relativistic approach applying PBE~+~U (U$_{4f}$~=~6~eV).
Two slightly different values of $m_\mathrm{s}$ for $++--$ configuration are present within single solution.
\vspace{2mm}
}
\centering
\begin{tabular}{ccccccc}
\hline \hline
		&\multicolumn{2}{c}{FM ($++++$)}&\multicolumn{2}{c}{AFM ($+-+-$)}&\multicolumn{2}{c}{AFM ($++--$)}\\
\hline       
axis   	& $m_\mathrm{s}$   & $m_\mathrm{l}$     & $m_\mathrm{s}$   & $m_\mathrm{l}$& $m_\mathrm{s}$   & $m_\mathrm{l}$     \\
\hline
$[001]$ & 1.001	& -0.563 & 1.003 & -0.561 & 1.000/1.004	& -0.562       \\
$[100]$	& 1.001	& -0.571 & 1.004 & -0.572 & 1.001/1.004	& -0.571       \\
\hline \hline
\end{tabular}
\end{table}
The calculated magnetic moments on Ce sites for considered colinear magnetic configurations are presented in Table~\ref{tab_mm_mag_configs}.

%
%
\begin{figure}[!t]
\centering
\includegraphics[clip, width = 0.9 \columnwidth]{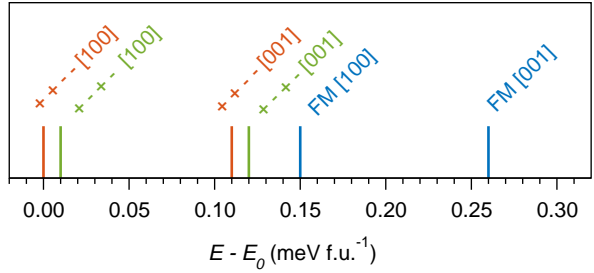}
\caption{\label{fig_de_mag_configs}The energy distance to the antiferromagnetic $++--$ [100] configuration ($E_0$).
The considered colinear magnetic configurations of \cecoge{} are based on the double unit cell model, see Fig.~\ref{fig_mag_stuct}.
Total energies are 
calculated with FPLO18 in fully-relativistic approach and applying PBE~+~U (U$_{4f}$~=~6~eV).
}
\end{figure}
The comparison of total energies, see Fig.~\ref{fig_de_mag_configs}, indicates that in each case the [100] magnetization direction is energetically more stable than the [001], which means that the magnetic moments prefer to order in $ab$ plane than along $c$ axis.
The similar orientation of magnetic moments in the ground state magnetic configuration has been suggested from experiment~\cite{ pecharsky1993unusual}.
However, in our model, we do not consider the small canting of magnetic moments suggested by the same experimental results~\cite{ pecharsky1993unusual}.
In all our cases the energy difference between [100] and [001] solutions is about 0.1~meV~f.u.$^{-1}$ (about 0.2~MJ~m$^{-3}$) which is comparable to the values of magnetocrystalline anisotropy obtained for semi-hard ferromagnets~\cite{golden_evolution_2018}.
A further comparison of all cases shows that the lower energies have been determined for antiferromagnetic solutions,
with the lowest energy being calculated for $++--$ [100] configuration, which we since now will consider as the ground state solution (from among the considered colinear cases).
Although the whole set of the obtained results indicates $++--$ solution as the ground state,
the energy distance between $++--$ [100] and $+-+-$ [100] solution is very small (about 0.01~meV~f.u.$^{-1}$) and lies at the limit of the accuracy we can get, 
so the final conclusion requires even more precise calculations than those carried out by us.

\end{document}